\documentstyle[12pt]{article}
\input amssym.def
\input amssym.tex

\newcommand{\qed}{\rule{3mm}{3mm}}
\newcommand{\eto}[1]{\mbox{$e^{\displaystyle #1}$}}
\newcommand{\nm}{\!-\!}
\newcommand{\np}{\!+\!}

\def\m@th{\mathsurround=0pt}
\mathchardef\bracell="0365 
\def\upbrall{$\m@th\bracell$}
\def\undertilde#1{\mathop{\vtop{\ialign{##\crcr
    $\hfil\displaystyle{#1}\hfil$\crcr
     \noalign
     {\kern1.5pt\nointerlineskip}
     \upbrall\crcr\noalign{\kern1pt
   }}}}\limits}

\newcommand{\g}{\mbox{\boldmath$g$}}
\newcommand{\G}{\mbox{\boldmath$G$}}

\newcommand{\rA}{{\rm A}}

\newcommand{\rH}{{\rm H}}

\newcommand{\rP}{{\rm P}}
\newcommand{\rR}{{\rm R}}

\newcommand{\rS}{{\rm S}}
\newcommand{\rW}{{\rm W}}

\newcommand{\bA}{{\bf A}}

\newcommand{\bL}{{\bf L}}
\newcommand{\bM}{{\bf M}}

\newcommand{\bS}{{\bf S}}

\newcommand{\ba}{{\bf a}}
\newcommand{\bb}{{\bf b}}

\newcommand{\bg}{{\bf g}}

\newcommand{\bu}{{\bf u}}
\newcommand{\bv}{{\bf v}}
\newcommand{\bw}{{\bf w}}

\newcommand{\mbA}{\mbox{\boldmath$A$}}
\newcommand{\mbB}{\mbox{\boldmath$B$}}
\newcommand{\mbC}{\mbox{\boldmath$C$}}
\newcommand{\mbD}{\mbox{\boldmath$D$}}

\newcommand{\mbL}{\mbox{\boldmath$L$}}
\newcommand{\mbP}{\mbox{\boldmath$P$}}

\newcommand{\mbU}{\mbox{\boldmath$U$}}
\newcommand{\mbV}{\mbox{\boldmath$V$}}
\newcommand{\mbW}{\mbox{\boldmath$W$}}

\newcommand{\mba}{\mbox{\boldmath$a$}}
\newcommand{\mbb}{\mbox{\boldmath$b$}}

\newcommand{\mbd}{\mbox{\boldmath$d$}}
\newcommand{\mbf}{\mbox{\boldmath$f$}}
\newcommand{\mbu}{\mbox{\boldmath$u$}}
\newcommand{\mbv}{\mbox{\boldmath$v$}}
\newcommand{\mbw}{\mbox{\boldmath$w$}}

\newcommand{\bphi}{\mbox{\boldmath$\phi$}}
\newcommand{\bpsi}{\mbox{\boldmath$\psi$}}
\newcommand{\bchi}{\mbox{\boldmath$\chi$}}

\newcommand{\cA}{{\cal A}}
\newcommand{\cB}{{\cal B}}

\newcommand{\cE}{{\cal E}}
\newcommand{\cF}{{\cal F}}

\newcommand{\cM}{{\cal M}}

\newcommand{\cR}{{\cal R}}

\newcommand{\cT}{{\cal T}}
\newcommand{\cV}{{\cal V}}

\newcommand{\cX}{{\cal X}}

\newcommand{\wa}{\widetilde{a}}
\newcommand{\wb}{\widetilde{b}}

\newcommand{\wv}{\widetilde{v}}
\newcommand{\wu}{\widetilde{u}}
\newcommand{\ww}{\widetilde{w}}

\newcommand{\wL}{\widetilde{L}}

\newcommand{\wU}{\widetilde{U}}
\newcommand{\wV}{\widetilde{V}}
\newcommand{\wW}{\widetilde{W}}

\newtheorem{theorem}{Theorem}[section]

\newtheorem{definition}[theorem]{Definition}

\begin{document}

\begin{titlepage}
\begin{center} {\LARGE What is the relativistic Volterra lattice?}
\end{center} 
\vspace{1.5cm}

\begin{center}{\large Yuri B. SURIS
\footnote{Centre for Complex Systems and Visualization, University of Bremen,
Universit\"atsallee 29, 28359 Bremen, Germany; {\it present address}: Technische
Universit\"at Berlin, Fachbereich Mathematik, SFB 288, Sekr. MA 8--5, 
Str. des 17. Juni 136, 10623 Berlin, Germany; 
e-mail: suris @ sfb288.math.tu-berlin.de}
and Orlando RAGNISCO \footnote{Dipartimento di Fisica, Universita di Roma Tre,
Via Vasca Navale 84, 00146 Roma, Italy;
e-mail: ragnisco @ amaldi.fis.uniroma3.it}}\end{center}
\vspace{1.5cm}

{\small{\bf Abstract.} We develop a systematic procedure of finding integrable
''relativistic'' (regular one--parameter) deformations for integrable lattice
systems. Our procedure is based on the integrable time discretizations and 
consists of three steps. First, for a given system one finds a local
discretization living in the same hierarchy. Second, one considers this
discretization as a particular Cauchy problem for a certain 2--dimensional
lattice equation, and then looks for another meaningful Cauchy problems,
which can be, in turn, interpreted as new discrete time systems.
Third, one has to identify integrable hierarchies to which these new 
discrete time systems belong. These novel hierarchies are called then
''relativistic'', the small time step $h$ playing the role of inverse 
speed of light. We apply this procedure to the Toda lattice (and
recover the well--known relativistic Toda lattice), as well as to the
Volterra lattice and a certain Bogoyavlensky lattice, for which the 
''relativistic'' deformations were not known previously.}

\end{titlepage}
%%%%%%%%%%%%%%%%%%%%%%%%%%%%%%%%%%%%%%%%%%%%%%%%%%%%%%%%%%%%%%%%%%%%%%%%%%
%\newpage
\setcounter{equation}{0}
\section{Introduction}\label{Introduction}

The theory of integrable differential--difference, or lattice, systems
is by now a well developed and well understood subject. Nevertheless,
some intriguing questions remain open, and the aim of this paper is to
close one of them: what is relativistic Volterra lattice? Let us start 
with the necessary background.

Certainly, two most celebrated and well studied integrable lattice systems
are the Toda lattice (TL),
\begin{equation}\label{TL}
\dot{b}_k=a_k-a_{k-1},\qquad \dot{a}_k=a_k(b_{k+1}-b_k)
\end{equation}
and the Volterra lattice (VL),
\begin{equation}\label{VL}
\dot{u}_k=u_k(v_k-v_{k-1}),\qquad \dot{v}_k=v_k(u_{k+1}-u_k)
\end{equation}
For readers more familiar with another forms of these systems, we recall
that the Newtonian form of the Toda lattice,
\begin{equation}\label{TL New}
\ddot{x}_k=\eto{x_{k+1}\nm x_k}-\eto{x_k\nm x_{k-1}}
\end{equation}
equivalent also to the Hamiltonian one,
\begin{equation}\label{TL Ham}
\dot{x}_k=p_k,\qquad \dot{p}_k=\eto{x_{k+1}\nm x_k}-\eto{x_k\nm x_{k-1}}
\end{equation}
is recovered from (\ref{TL}), if the variables $a_k$, $b_k$ are parametrized
according to Manakov--Flaschka formulas
\begin{equation}\label{TL l par}
b_k=p_k,\qquad a_k=\eto{x_{k+1}\nm x_k}
\end{equation}
while the more convenient form of the Volterra lattice,
\begin{equation}\label{VL in a}
\dot{a}_k=a_k(a_{k+1}-a_{k-1})
\end{equation}
arises from (\ref{VL}) upon the re-naming
\begin{equation}
u_k=a_{2k-1},\qquad v_k=a_{2k}
\end{equation}

These two lattice systems are connected to one another in two different ways.
On the one hand, the Volterra lattice (\ref{VL in a}) is a restriction to the 
manifold $b_k=0$ of the second flow of the Toda hierarchy. On the other hand
(and this connection will be of a primary interest for us here) the flows
(\ref{TL}) and (\ref{VL}) are connected by a Miura map, or, better, by two
different Miura maps $\cM_{1,2}:(u,v)\mapsto (a,b)$:
\begin{equation}\label{Miuras}
\cM_1:\left\{\begin{array}{l} b_k=u_k+v_{k-1}\\ a_k=u_kv_k\end{array}\right.
\qquad
\cM_2:\left\{\begin{array}{l} b_k=u_k+v_k\\ a_k=u_{k+1}v_k\end{array}\right.
\end{equation}

Some time ago a remarkable discovery was made by Ruijsenaars \cite{R}
in the area
of integrable lattice systems: he found a relativistic generalization of TL
(RTL). The corresponding system may be viewed as a regular deformation of TL:
\begin{equation}\label{RTL}
\dot{\mbd}_k=(1+h \mbd_k)(\mba_k-\mba_{k-1}), \qquad
\dot{\mba}_k=\mba_k(\mbd_{k+1}-\mbd_k+h \mba_{k+1}-h \mba_{k-1})
\end{equation}
(the reason for choosing here $\mbd$ instead of $\mbb$ will become clear in the
main text). Under the parametrization
\begin{equation}\label{RTL q par}
\mbd_k=\left(\eto{h  p_k}-1\right)/h ,\qquad 
\mba_k=\eto{x_{k+1}\nm x_k+h  p_k}
\end{equation} 
(which is a regular deformation of (\ref{TL l par})) the equations of motion
(\ref{RTL}) may be presented as
\begin{equation}\label{RTL Ham}
\left\{\begin{array}{l}
\dot{p}_k =\displaystyle\eto{x_{k+1}\nm x_k\np h  p_k}-
\displaystyle\eto{x_k\nm x_{k-1}\np h  p_{k-1}}\\ \\
1+h \dot{x}_k =
\eto{h  p_k}\left(1+h ^2\eto{x_{k+1}\nm x_k}
\right)\end{array}\right.
\end{equation}
which implies also Newtonian equations of motion
\begin{equation}\label{RTL New}
\ddot{x}_k=(1+h \dot{x}_k)(1+h \dot{x}_{k+1}) \displaystyle\frac
{\eto{x_{k+1}\nm x_k}}{1+h ^2\eto{x_{k+1}\nm x_k}}-
(1+h \dot{x}_{k-1})(1+h \dot{x}_k) \displaystyle\frac
{\eto{x_k\nm x_{k-1}}}{1+h ^2\eto{x_k\nm x_{k-1}}}
\end{equation}
Mathematical structures related to RTL, including Lax representations, 
multi--Hamiltonian structure, and so on, were further investigated in
\cite{BR}, \cite{ZTOF}, \cite{S1}, \cite{S2}. A general approach to 
constructing ''relativistic'' generalizations of integrable lattice
systems, applicable to the whole lattice KP hierarchy, was proposed in
\cite{GK1}.

Paradoxically, up to now nobody seems to know what relativistic Volterra 
lattice is. We propose here an answer to this question. 
Let us stress that we are not concerned here with relativistic invariance.
Instead, our aim is to construct integrable lattice systems, which
are regular one--parameter deformations of VL and are Miura--related to
RTL in the same manner as VL is related to TL. Needless to say that the
''relativistic Miura maps'' we are looking for have to be regular deformations 
of the standard ones. 

Surprisingly, these Miura maps can be choosen to be {\it identical} with
the nonrelativistic ones. There exist two different systems that are sent to 
the RTL by either of two Miura maps (\ref{Miuras}), $\cM_{1,2}(\mbu,\mbv)=
(\mba,\mbd)$, so that ''the relativistic Volterra lattice'' actually splits 
into two systems:
\begin{equation}\label{RVL+}
\left\{\begin{array}{c}
\dot{\mbu}_k=\mbu_k(\mbv_k-\mbv_{k-1}+
h \mbu_k\mbv_k-h \mbu_{k-1}\mbv_{k-1})\\ \\
\dot{\mbv}_k=\mbv_k(\mbu_{k+1}-\mbu_k+
h \mbu_{k+1}\mbv_{k+1}-h \mbu_k\mbv_k)
\end{array}\right.
\end{equation}
and
\begin{equation}\label{RVL-}
\left\{\begin{array}{c}
\dot{\mbu}_k=\mbu_k(\mbv_k-\mbv_{k-1}+
h \mbu_{k+1}\mbv_k-h \mbu_k\mbv_{k-1})\\ \\
\dot{\mbv}_k=\mbv_k(\mbu_{k+1}-\mbu_k+
h \mbu_{k+1}\mbv_k-h \mbu_k\mbv_{k-1})
\end{array}\right.
\end{equation}

We suspect that our way to finding these systems might be more significant than 
the systems themselves. Namely, our route is through the theory of {\it 
integrable time discretizations}. It was discovered in \cite{S1} that certain 
integrable discretization of the usual TL, 
\begin{equation}\label{edTL int}
\widetilde{\mbb}_k=\mbb_k+h(\mba_k-\mba_{k-1}), \qquad
\widetilde{\mba}_k(1+h\widetilde{\mbb}_k)=\mba_k(1+h\widetilde{\mbb}_{k+1})
\end{equation}
shares the integrals of motion and the invariant Poisson structure with RTL
(upon the change of variables $\mbb_k=\mbd_k+h\mba_{k-1}$). In other words, 
this discretization belongs to the RTL hierarchy. In \cite{PGR}, \cite{S3} 
it was noticed  that this discretization is connected to other one, belonging 
to the TL hierarchy:
\begin{equation}\label{dTL int}
\widetilde{\bb}_k=\bb_k+h(\ba_k-\widetilde{\ba}_{k-1}), \qquad
\widetilde{\ba}_k(1+h\widetilde{\bb}_k)=\ba_k(1+h\bb_{k+1})
\end{equation}
These two 1+1-dimensional discrete systems (with one
discrete space coordinate and one discrete time) may be seen as resuling from
one and the same 2-dimensional discrete system (living on a 2-dimensional 
lattice) by posing an {\it initial value problem} in two different ways.
In other words, the Cauchy data for two discretizations are prescribed on two
different ''discrete curves'' on the 2-dimensional lattice. Being very
close on the 2--dimensional lattice, the equations (\ref{edTL int}), 
(\ref{dTL int}) have nevertheless very different properties, such as invariant 
Poisson brackets, integrals of motion, Lax matrices, etc. -- all in one, they
belong to different hierarchies.

Thus, our strategy to finding the ''relativistic'' deformations for continuous
time lattice systems (with one discrete space coordinate) may be described as
follows.

First, for a given integrable lattice system, one constructs integrable 
time discretization belonging to the same hierarchy. Such time discretizations 
appeared first in \cite{AL}, \cite{GK2}, and a systematic procedure was
developed in \cite{S3}, \cite{S4}. As a rule, this approach results in nonlocal 
equations of motion. However, these nonlocal discrete time equations of motion 
may be brought into a local form  with the help of the so called localizing
changes of variables, which were found in \cite{S6} for a large set of examples.

Second, and this is a crucial step in finding new hierarchies, one 
considers the resulting discrete time system as a system on a
two--dimensional lattice, and tries to find new meaningful initial
value problems for this two--dimensional lattice. This is close in spirit
to the constructions in \cite{PNC}, \cite{NPCQ}, \cite{PN}. The resulting 
discrete time systems belong to integrable hierachies distinct from the 
original one (this fact is not stressed in the papers just mentioned).

As the third and final step, one has to identify these novel hierarchies,
in particular, to find integrals of motion, invariant Poisson structures,
and higher flows.

In this paper we first recall how this program could be realized to find 
RTL (although the actual way to discovering this system was quite different),
and then demonstrate how the relativistic Volterra lattice may be derived.
In particular, we show that (\ref{RVL+}) and (\ref{RVL-}) are the 
simplest flows of the hierarchies, to which the following two explicit 
discretizations of VL belong:
\begin{equation}\label{edVL+ int}
\widetilde{\mbu}_k(1+h\mbv_{k-1})=\mbu_k(1+h\mbv_k), \qquad
\widetilde{\mbv}_k(1+h\widetilde{\mbu}_k)=\mbv_k(1+h\widetilde{\mbu}_{k+1})
\end{equation}
and
\begin{equation}\label{edVL- int}
\widetilde{\mbu}_k(1+h\widetilde{\mbv}_{k-1})=\mbu_k(1+h\widetilde{\mbv}_k), 
\qquad \widetilde{\mbv}_k(1+h\mbu_k)=\mbv_k(1+h\mbu_{k+1})
\end{equation}
respectively. One can see that the corresponding constructions may be described
as a factorization of RTL, in a complete analogy with the nonrelativistic case.
Finally, we show how to generalize these results to some of the Bogoyavlensky 
lattices.

\setcounter{equation}{0}
\section{General framework}\label{Sect framework}
\subsection{Lax equations and representations}
Our approach to integrable lattice systems is based on the notion of Lax
representations. We consider Lax equations of one of the following types:
\begin{equation}\label{Lax in recipe}
\dot{L}=\Big[\,L,\pi_+(f(L))\,\Big]=-\Big[\,L,\pi_-(f(L))\,\Big]
\end{equation}
or
\begin{equation}\label{Lax triads in recipe}
\dot{L}_j\; =\; L_j\cdot\pi_+\Big(f(T_{j-1})\Big)-\pi_+\Big(f(T_j)\Big)\cdot L_j
\; = \; -\,L_j\cdot\pi_-\Big(f(T_{j-1})\Big)+\pi_-\Big(f(T_j)\Big)\cdot L_j
\end{equation}
Let us explain the notations. 

Let $\g$ be an associative algebra. One can introduce in $\g$ 
the structure of Lie algebra in a standard way. Let $\g_+$, $\g_-$ be
two subalgebras such that as a vector space $\g$ is a direct sum
$\g=\g_+\oplus \g_-$. Denote by $\pi_{\pm}:\g\mapsto\g_{\pm}$ the corresponding
projections. Finally, let $f:\g\mapsto\g$ be an ${\rm Ad}$--covariant
function on $\g$, and let $L$ stand for a generic element of $\g$. 
Then (\ref{Lax in recipe}) is a certain differential equation on $\g$. 

Further, let $\bg=\bigotimes_{j=1}^m\g$ be a direct product of $m$ copies
of the algebra $\g$. A generic element of $\bg$ is denoted by
$\bL=(L_1,\ldots,L_m)$. We use also the notation
\begin{equation}\label{Tj}
T_j=T_j(\bL)=L_j\cdot\ldots\cdot L_1\cdot L_m\cdot\ldots\cdot L_{j+1}
\end{equation}
Then (\ref{Lax triads in recipe}) is a certain differential equation on $\bg$.
Such equations are sometimes called {\it Lax triads}.

One says that (\ref{Lax in recipe}), resp. (\ref{Lax triads
in recipe}), is a Lax representation of a Hamiltonian flow 
\begin{equation}\label{Ham syst}
\dot{x}=\{H,x\}
\end{equation} 
on a Poisson manifold $\Big(\cX,\{\cdot,\cdot\}\Big)$, if there exists a map 
$L:\cX\mapsto\g$ (resp. $\bL:\cX\mapsto\bg$)
such that the former equations of motion are equivalent to the latter ones.
Let us stress that when considering equations (\ref{Lax in recipe}), resp. 
(\ref{Lax triads in recipe}) in the role of Lax representation, the letter 
$L$ (resp. $\bL$) does not stand for a generic element of the corresponding 
algebra any more; rather, it represents the elements of the images of the maps 
$L:\cX\mapsto\g$ and $\bL:\cX\mapsto\bg$, correspondingly. The elements $L(x)$, 
resp. $\bL(x)$ (and the map $L$, resp. $\bL$, itself) are called {\it Lax 
matrices}. 

\subsection{$r$--matrix Poisson brackets}

Recall that there exist several constructions of Poisson brackets on associative
algebras implying the Lax form of Hamiltonian equations of motion. We recall
here some of them.

Suppose that $\g$ carries a nondegenerate scalar product $\langle\cdot,
\cdot\rangle$, bi--invariant with respect to the multiplication in $\g$. Let 
$\rR$ be a linear operator on $\g$.

\begin{definition} {\rm\cite{STS}}  A {\rm linear $r$--matrix bracket} on 
$\g$ corresponding to the operator $\rR$ is defined by:
\begin{equation}\label{lin br}
\{\varphi,\psi\}_1(L)=\frac{1}{2}\Big\langle\, 
[\rR(\nabla \varphi(L)),\nabla \psi(L)]+[\nabla \varphi(L),\rR(\nabla \psi(L))]
\,,\,L\,\Big\rangle
\end{equation}
If this is indeed a Poisson bracket, it will denoted by ${\rm PB}_1(\rR)$. 
\end{definition}
\begin{theorem} {\rm\cite{STS}} A sufficient condition for {\rm(\ref{lin br})}
to define a Poisson bracket is given by the {\rm modified Yang--Baxter equation} for the operator $\rR$, 
${\rm mYB}(\rR;\alpha)$, which reads
\begin{equation}\label{mYB}
[\rR(u),\rR(v)]-\rR\Big([\rR(u),v]+[u,\rR(v)]\Big)=-\alpha\,[u,v]
\qquad\forall u,v\in\g
\end{equation}
\end{theorem}

Now let $\rA_1$, $\rA_2$, $\rS$ be three linear operators on $\g$, $\rA_1$ 
and $\rA_2$ being skew--symmetric:
\begin{equation}\label{skew}
\rA_1^*=-\rA_1,\qquad \rA_2^*=-\rA_2
\end{equation}

\begin{definition} {\rm\cite{S2}} A {\rm quadratic $r$--matrix bracket} on
$\g$ corresponding to the triple $\rA_1$, $\rA_2$, $\rS$ is defined by:
\begin{eqnarray}
\{\varphi,\psi\}_2(L) & = & 
\frac{1}{2}\,\Big\langle\rA_1(d\,'\varphi(L)),\,d\,'\psi(L)\,\Big\rangle
-\frac{1}{2}\,\Big\langle\rA_2(d\varphi(L)),\,d\psi(L)\,\Big\rangle 
\nonumber\\ \nonumber\\
                      &   & 
+\frac{1}{2}\,\Big\langle\rS(d\varphi(L)),\,d\,'\psi(L)\,\Big\rangle
-\frac{1}{2}\,\Big\langle\rS^*(d\,'\varphi(L)),\,d\psi(L)\,\Big\rangle 
\label{q br}
\end{eqnarray}
where we denote for brevity
\begin{equation}\label{dd'}
d\varphi(L)=L\cdot\nabla\varphi(L),\qquad 
d\,'\varphi(L)=\nabla\varphi(L)\cdot L
\end{equation}
If this expression indeed defines a Poisson bracket, we shall denote 
it by ${\rm PB}_2(\rA_1,\rA_2,\rS)$.
\end{definition}
In what follows we shall usually suppose the following condition to be satisfied:
\begin{equation}\label{sum}
\rA_1+\rS=\rA_2+\rS^*=\rR
\end{equation}
Then a linearization of ${\rm PB}_2(\rA_1,\rA_2,\rS)$ in the unit element of
$\g$ coincides with ${\rm PB}_1(\rR)$, and we call the former a {\it
quadratization} of the latter.
\begin{theorem} {\rm\cite{S2}} A sufficient condition for {\rm(\ref{q br})} 
to be a Poisson bracket is given by the equations {\rm(\ref{sum})} and
\begin{equation}\label{q br suf1}
{\rm mYB}(\rR;\alpha),\quad {\rm mYB}(\rA_1;\alpha),\quad 
{\rm mYB}(\rA_2;\alpha)
\end{equation}
\end{theorem}

One of the most important properties of the $r$--matrix brackets is the 
following one.

\begin{theorem} ${\rm Ad}$--invariant functions on $\g$ are in involution with
respect to the bracket ${\rm PB}_1(\rR)$ and with respect to its 
quadratizations ${\rm PB}_2(\rA_1,\rA_2,\rS)$. The Hamiltonian equations 
of motion on $\g$ corresponding to an ${\rm Ad}$--invariant Hamilton function 
$\varphi$, have the Lax form
\begin{equation}
\dot{L}=\frac{1}{2}\Big[L,\rR(f(L))\Big]
\end{equation}
where $f(L)=\nabla\varphi(L)$ for the linear $r$--matrix bracket, and 
$f(L)=d\varphi(L)$ for its quadratizations.
\end{theorem}

Quadratic $r$--matrix brackets have interesting and important features when
considered on a ''big'' algebra $\bg=\bigotimes_{j=1}^m\g$. This algebra
carries a (nondegenerate, bi--invariant) scalar product 
\[
\langle\langle\bL,\bM\rangle\rangle=\sum_{k=1}^m\langle L_k,M_k\rangle
\]
Working with linear operators on $\bg$, we use the following natural notations.
Let $\bA:\bg\mapsto\bg$ be a linear operator, let 
$\Big(\bA(\bL)\Big)_i$ be the $i$th component of $\bA(\bL)$; then we set
\begin{equation}\label{A i}
\Big(\bA(\bL)\Big)_i=\sum_{j=1}^m (\bA)_{ij}(L_j)
\end{equation}
For a smooth function $\Phi(\bL)$ on $\bg$ we also denote by $\nabla_j\Phi$, 
$d_j\Phi$, $d\,'_{\!j}\Phi$ the $j$th components of the corresponding objects.

Now let $\bA_1, \bA_2, \bS$ be linear operators on $\bg$ satisfying conditions
analogous to (\ref{skew}) and to (\ref{q br suf1}). One has, obviously:
\[
\Big((\bA_1)_{ij}\Big)^*=-(\bA_1)_{ji}, \qquad 
\Big((\bA_2)_{ij}\Big)^*=-(\bA_2)_{ji}, \qquad
(\bS_{ij})^*=(\bS^*)_{ji}
\]
Then one can define the bracket ${\rm PB}_2(\bA_1,\bA_2,\bS)$ on $\bg$. In
components it reads:
\begin{eqnarray}\label{big br in components}
\{\Phi,\Psi\}_2(\bu) & = &
\frac{1}{2}\sum_{i,j=1}^m\Big\langle (\bA_1)_{ij}(d\,'_{\!j}\Phi),
d\,'_{\!i}\Psi\Big\rangle-
\frac{1}{2}\sum_{i,j=1}^m\Big\langle (\bA_2)_{ij}(d_j\Phi),
d_i\Psi\Big\rangle\nonumber\\
 & & +\frac{1}{2}\sum_{i,j=1}^m\Big\langle \bS_{ij}(d_j\Phi),
d\,'_{\!i}\Psi\Big\rangle
-\frac{1}{2}\sum_{i,j=1}^m\Big\langle (\bS^*)_{ij}(d\,'_{\!j}\Phi),
d_i\Psi\Big\rangle
\end{eqnarray}
\begin{theorem}\label{monodromy} {\rm\cite{S5}}
Let $\bg$ be equipped with the Poisson bracket ${\rm PB}_2(\bA_1,\bA_2,\bS)$.
Suppose that the following relations hold:
\[
(\bA_1)_{i+1,j+1}=-(\bS)_{i+1,j}=(\bS^*)_{i,j+1}=-(\bA_2)_{i,j}\quad
{\rm for}\quad i\neq j
\]
\[
(\bA_1)_{j+1,j+1}-(\bA_2)_{j,j}+(\bS)_{j+1,j}-(\bS^*)_{j,j+1}=0\quad
{\rm for\;\;all}\quad 1\le j\le m
\]
Then each map $T_j:\bg\mapsto\g$ {\rm(\ref{Tj})} is Poisson, if the target 
space $\g$ is equipped with the Poisson bracket
\[
{\rm PB}_2
\Big((\bA_1)_{j+1,j+1},(\bA_2)_{j,j},(\bS)_{j+1,j}\Big)
\] 
Hamilton function of the form $\Phi(\bL)=\varphi(L_m\cdot ... \cdot L_1)$, 
where $\varphi$ is an ${\rm Ad}$--invariant function on $\g$,
generates Hamiltonian equations of motion on $\bg$ having the Lax form:
\begin{equation}
\dot{L}_j=L_j\,{\cal B}_{j-1}-{\cal B}_j L_j,\qquad
{\cal B}_j=\frac{1}{2}\,\rR_j(d\varphi(T_j))
\end{equation}
where 
\[
\rR_j=(\bA_1)_{j+1,j+1}+(\bS)_{j+1,j}=(\bA_2)_{j,j}+(\bS^*)_{j,j+1}
\]
(In all the formulas the subscripts should be taken {\rm (mod $m$)}).
\end{theorem}

We have discussed above the $r$--matrix origin of Lax equations. If one is 
concerned with a Lax representation of a Hamiltonian flow (\ref{Ham syst})
on a Poisson manifold $\Big(\cX,\{\cdot,\cdot\}\Big)$, then finding an 
$r$--matrix interpretation for it consists of finding an $r$--matrix
bracket on $\g$ (or on $\bg$) such that the Lax matrix map $L:\cX\mapsto\g$ 
(resp. $\bL:\cX\mapsto\bg$) is a Poisson map.

\subsection{Factorizations and integrable discretization}\label{Sect recipe}

A further remarkable feature of the  equations (\ref{Lax in recipe}) and 
(\ref{Lax triads in recipe}) is a possibility to solve them
explicitly in terms of a certain factorization problem 
in the Lie group $\G$ corresponding to $\g$ \cite{Sy}, \cite{STS}, 
\cite{RSTS}. (Actually, this can be done even in a more general situation 
of hierarchies governed by $R$--operators satisfying the modified 
Yang--Baxter equation, see \cite{RSTS}). The factorization problem 
is described by the equation
\begin{equation}\label{fact problem}
U=\Pi_+(U)\cdot\Pi_-(U),\quad U\in\G,\quad\Pi_{\pm}(U)\in \G_{\pm}
\end{equation}
where $\G_{\pm}$ are two subgroups of $\G$ with the Lie algebras $\g_{\pm}$, 
respectively. This problem has a unique solution in a certain neighbourhood
of the group unit. 

In the situations we consider in the sequel $\G$ will be a matrix group, 
and we write the adjoint action of the group elements on $\g$ as a conjugation 
by the corresponding matrices. In this context we write 
$\Pi_{\pm}^{-1}(U)$ for $(\Pi_{\pm}(U))^{-1}$. Correspondingly, 
we call ${\rm Ad}$--covariant functions $\g\mapsto\g$ also ''conjugation
covariant''. This notation has an additional advantage of being applicable
also to functions $\g\mapsto\G$.

Based on the above mentioned explicit solution, the following recipe for
integrable discretization was formulated in \cite{S3}, \cite{S4}. In all
the difference equations below we use the tilde to denote the discrete time
shift, so that, for example, $\wL=L(t+h)$, if $L=L(t)$, $t\in h{\Bbb Z}$.
\vspace{2mm}

{\bf Recipe.} {\it Suppose you are looking for an integrable discretization
of an integrable system {\rm(\ref{Ham syst})} allowing a Lax representation 
of the form {\rm(\ref{Lax in recipe})}. Then as a solution of your task you may 
take the difference equation 
\begin{equation}\label{dLax in recipe}
\wL=\Pi_+^{-1}(F(L))\cdot L\cdot\Pi_+(F(L))=\Pi_-(F(L))\cdot L\cdot
\Pi_-^{-1}(F(L))
\end{equation}
with the same Lax matrix $L$ and some conjugation covariant function
$F:\g\mapsto\G$ such that
\[
F(L)=I+hf(L)+O(h^2)
\]
Analogously, if your system has a Lax representation of the form {\rm(\ref{Lax 
triads in recipe})} on the algebra $\bg$, then you may 
take as its integrable discretization the difference Lax equation
\begin{equation}\label{dLax triads in recipe}
\wL_j \;= \; \Pi_+^{-1}\!\left(F(T_j)\right)\cdot
L_j\cdot\Pi_+\!\left(F(T_{j-1})\right)
\; = \; \Pi_-\!\left(F(T_j)\right)\cdot
L_j\cdot\Pi_-^{-1}\!\left(F(T_{j-1})\right)
\end{equation}
with $F$ as above.}
\vspace{2mm}

Obviously, by construction, the discretizations obtained in this way share 
the Lax matrix and therefore the integrals of motion with their underlying 
continuous time systems. Moreover, they share also the invariant Poisson
brackets. Indeed, the above mentioned factorization theorems imply that the 
maps (\ref{dLax in recipe}), (\ref{dLax triads in recipe}) are the 
time $h$ shifts along the trajectories of the corresponding flows 
(\ref{Lax in recipe}), (\ref{Lax triads in recipe}) with
\[
\mbf(L)=h^{-1}\log(F(L))=f(L)+O(h).
\]
This, in turn, implies that if all flows of the hierarchy (\ref{Lax in recipe}) 
[resp. (\ref{Lax triads in recipe})] are Hamiltonian with respect to a certain
Poisson bracket, then our discretizations are Poisson maps with respect to
this bracket. In particular, if the Lax matrices $L$ [resp. $\bL$] form a 
Poisson submanifold for some $r$--matrix bracket, then this submanifold is left 
invariant by the corresponding Poisson map (\ref{dLax in recipe}) [resp. 
(\ref{dLax triads in recipe})]. We shall say that our recipe yields 
discretizations living in the same hierarchies as the underlying continuous 
time systems.

\subsection{Basic algebras and operators}

In what follows we fix an algebra $\g$ suited to describe various lattice 
systems with periodic boundary conditions, as well as certain operators 
taking part in the corresponding $r$--matrix constructions. So, starting
from this point the symbols $\g$, $\g_{\pm}$, $\G$, $\G_{\pm}$,
$\pi_{\pm}$, $\Pi_{\pm}$, $\rR$, $\rA_1$, $\rA_2$, $\rS$ will carry only
the following meanings.

The algebra $\g$ is a {\it twisted loop algebra} over $gl(N)$, consisting  
of Laurent polynomials $L(\lambda)$ with coefficients from $gl(N)$ 
and a natural commutator $[u\lambda^j,v\lambda^k]=[u,v]\lambda^{j+k}$,
satisfying an additional condition
\[
\Omega L(\lambda)\Omega^{-1}=L(\omega\lambda),\;\;{\rm where}\;\;
\Omega={\rm diag}(1,\omega,\ldots,\omega^{N-1}),\;\;\omega=\exp(2\pi i/N)
\]
In other words, elements of $\g$ have the following structure:
\begin{equation}\label{L}
L(\lambda)=\sum_{p}\ell^{(p)}\cE^p
\end{equation}
In this formula $\ell^{(p)}={\rm diag}\left(\ell_1^{(p)},\ldots,\ell_N^{(p)}
\right)$ 
are diagonal matrices, and
\[
\cE=\lambda\sum_{k=1}^N E_{k+1,k},\qquad 
\cE^{-1}=\lambda^{-1}\sum_{k=1}^N E_{k,k+1}
\]
are the {\it shift matrices} (here and below $E_{jk}$ stands for the matrix 
whose only nonzero entry is on the intersection of the $j$th row and the $k$th 
column and is equal to 1; we set $E_{N+1,N}=E_{1,N}$, $E_{N,N+1}=E_{N,1}$, and
in general understand all subscripts (mod $N$)).

The nondegenerate bi--invariant scalar product on $\g$ is chosen as 
\begin{equation}
\langle L(\lambda),\,M(\lambda)\rangle={\rm tr}(L(\lambda)\cdot M(\lambda))_0
\end{equation}
the subscript 0 denoting the free term of the formal Laurent series. This 
scalar product allows to identify $\g^*$ with $\g$.

The two subalgebras $\g_+$, $\g_-$ such that $\g=\g_+\oplus\g_-$ are choosen as
\begin{equation}
\g_+=\left\{\sum_{p\ge 0}\ell^{(p)}\cE^p\right\},\qquad 
\g_-=\left\{\sum_{p<0}\ell^{(p)}\cE^p\right\}
\end{equation}
The corresponding decomposition $L=\pi_+(L)+\pi_-(L)$ of an arbitrary element
$L\in\g$ will be called the {\it generalized LU decomposition}.

We denote also by $\g_0$ the commutative subalgebra of $\g$ consisting of
$\lambda$--independent diagonal matrices (i.e. of matrices $\ell^{(0)}\cE^0$).
The linear operator on $\g$ assigning to each element $L(\lambda)$ (\ref{L}) 
its free term $\ell^{(0)}$ will be denoted by $\rP_0$.

The group $\G$ corresponding to the twisted loop algebra $\g$ i
is a {\it twisted loop group}, consisting of $GL(N)$--valued functions 
$U(\lambda)$ of the complex parameter $\lambda$, regular in 
${\Bbb C}P^1\backslash\{0,\infty\}$ and satisfying $\Omega U(\lambda)\Omega^{-1}
=U(\omega\lambda)$. Its subgroups
$\G_+$ and $\G_-$ corresponding to the Lie algebras $\g_+$ and $\g_-$,  
are singled out by the following conditions: the elements $U(\lambda)\in\G_+$ 
are regular in the neighbourhood of $\lambda=0$, and the elements
$U(\lambda)\in\G_-$ are regular in the neighbourhood of $\lambda=\infty$ and  
take in this point the value $U(\infty)=I$.
We call the corresponding $\Pi_+\Pi_-$ factorization the {\it generalized
$LU$ factorization}. 

The basic operator governing the hierarchies of Lax equations, is:
\begin{equation}\label{R}
\rR=\pi_+-\pi_-
\end{equation}
Denote by $\rR_0$, $\rP_0$ its skew--symmetric and its symmetric parts, 
respectively:
\[
\rR_0=(\rR-\rR^*)/2, \qquad \rP_0=(\rR+\rR^*)/2
\]
Obviously,  this definition of $\rP_0$ is consistent with the previous one.

Let the skew--symmetric operator $\rW$ act on $\g_0$ according to
\[
\rW(E_{jj})=\sum_{k<j}E_{kk}-\sum_{k>j}E_{kk}
\]
and on the rest of $\g$ according to $\rW=\rW\circ\rP_0$. Finally, define:
\begin{equation}\label{AS}
\rA_1=\rR_0+\rW,\quad \rA_2=\rR_0-\rW,\quad \rS=\rP_0-\rW,\quad
\rS^*=\rP_0+\rW
\end{equation}
These operators will be basic building blocks in the quadratic $r$--matrix 
brackets appearing below.

%%%%%%%%%%%%%%%%%%%%%%%%%%%%%%%%%%%%%%%%%%%%%%%%%%%%%%%%%%%%%%%%%%%%%%%%%

\setcounter{equation}{0}
\section{Reminding the Toda lattice case}
\subsection{TL}

We consider the equations of motion of TL (\ref{TL}) under periodic boundary 
conditions: all subscripts are taken (mod $N$). 
The phase space of this system is
\begin{equation}\label{TL phase sp}
\cT={\Bbb R}^{2N}(b_1,a_1,\ldots,b_N,a_N)
\end{equation}
(recall that $a_0\equiv a_N$, $b_{N+1}\equiv b_1$).

There exist three compatible local Poisson brackets on $\cT$ 
such that the system TL is Hamiltonian with respect to each one of them, see
\cite{K1}. 
We adopt once and forever the following conventions: the Poisson 
brackets will be defined by writing down {\it all nonvanishing} brackets between 
the coordinate functions; the indices in the corresponding formulas are
taken (mod $N$).

The ''linear'' Poisson structure on $\cT$ is defined by the brackets
\begin{equation}\label{TL l br}
\{b_k,a_k\}_1=-a_k, \qquad \{a_k,b_{k+1}\}_1=-a_k 
\end{equation}
the corresponding Hamilton function for the flow TL is given by:
\begin{equation}\label{TL H2}
\rH_2(a,b)=\frac{1}{2}\sum_{k=1}^N b_k^2+\sum_{k=1}^{N}a_k
\end{equation}
The ''quadratic'' Poisson structure has the following definition:
\begin{equation}\label{TL q br}
\begin{array}{cclcccl}
\{b_k,a_k\}_2 & = & -a_kb_k,&          \quad & 
\{a_k,b_{k+1}\}_2 & = & -a_kb_{k+1}  \\ \\
\{b_k,b_{k+1}\}_2 & = & -a_k,& \quad & 
\{a_k,a_{k+1}\}_2 & = & -a_{k+1}a_k 
\end{array}
\end{equation}
The Hamilton function generating TL in this bracket is:
\begin{equation}\label{TL H1}
\rH_1(a,b)=\sum_{k=1}^N b_k
\end{equation}
Finally, the ''cubic'' bracket on $\cT$ is given by the relations
\begin{equation}\label{TL c br}
\begin{array}{cclcccl}
\{b_k,a_k\}_3     & = & -a_k(b_k^2+a_k),     & \quad &
\{a_k,b_{k+1}\}_3 & = & -a_k(b_{k+1}^2+a_k), \\ \\
\{b_k,b_{k+1}\}_3 & = & -a_k(b_k+b_{k+1}),   & \quad &
\{a_k,a_{k+1}\}_3 & = & -2a_ka_{k+1}b_{k+1}, \\ \\
\{b_k,a_{k+1}\}_3 & = & -a_ka_{k+1},         & \quad & 
\{a_k,b_{k+2}\}_3 & = & -a_ka_{k+1}
\end{array}
\end{equation}
The corresponding Hamilton function of the flow TL is:
\begin{equation}\label{TL H0}
\rH_0(a,b)=\frac{1}{2}\sum_{k=1}^N \log(a_k)
\end{equation}

The Lax representation of the Toda lattice \cite{F}, \cite{M} lives in the 
algebra $\g$ introduced in the previous section. We shall work with the
Lax matrix
\begin{equation}\label{TL T}
L(a,b,\lambda)=\lambda\sum_{k=1}^{N} E_{k+1,k}+\sum_{k=1}^N b_kE_{k,k}+
\lambda^{-1}\sum_{k=1}^{N} a_kE_{k,k+1}=\cE+b+a\cE^{-1}
\end{equation}
where diagonal matrices $a={\rm diag}(a_1,\ldots,a_N)$, 
$b={\rm diag}(b_1,\ldots,b_N)$ are introduced.

The equations of motion {\rm (\ref{TL})} are equivalent to the Lax equations
\begin{equation}\label{TL Lax}
\dot{L}=[L,B]=-[L,A]
\end{equation}
with
\begin{eqnarray}
B & = & \pi_+(L)\;=\;
\lambda\sum_{k=1}^{N} E_{k+1,k}+\sum_{k=1}^Nb_kE_{k,k}\;=\;\cE+b
\label{TL B+}\\
A & = & \pi_-(L)\;=\;\lambda^{-1}\sum_{k=1}^{N} a_kE_{k,k+1}\;=\;a\cE^{-1}
\label{TL B-}
\end{eqnarray}
where $\pi_{\pm}\,:\,\g\mapsto\g_{\pm}$ are the projections to the subalgebras
$\g_{\pm}$.

Spectral invariants of the Lax matrix $L(a,b,\lambda)$ serve as integrals of 
motion of this system. Note that all Hamilton functions in different 
Hamiltonian formulations belong to these spectral invariants. For instance,
\[
\rH_2(a,b)=\frac{1}{2}\Big({\rm tr}\, L^2(a,b,\lambda)\Big)_0,\qquad
\rH_1(a,b)=\Big({\rm tr}\, L(a,b,\lambda)\Big)_0
\]
where the subscript ''0'' is used to denote the free term of the corresponding
Laurent series.
 
{\it All} spectral invariants turn out to be in involution with respect to
each of the Poisson brackets (\ref{TL l br}), (\ref{TL q br}), (\ref{TL c br}). 
Most directly it follows from the $r$--matrix interpretation of the Lax 
equation (\ref{TL Lax}).
\begin{theorem} The Lax matrix map $L(a,b,\lambda):\cT\mapsto\g$ is Poisson,
if $\cT$ carries $\{\cdot,\cdot\}_1$ and $\g$ carries ${\rm PB}_1(\rR)$, and
also if $\cT$ carries $\{\cdot,\cdot\}_2$ and $\g$ carries 
${\rm PB}_2(\rA_1,\rA_2,\rS)$.
\end{theorem}
The first statement is from \cite{AM}, the second one -- from \cite{S2}.
For other versions of such statements (including the $r$--matrix interpretation
of the cubic bracket) see \cite{DLT1}, \cite{OR}, \cite{MP}. 

\subsection{TL $\to$ dTL}

In order to find an integrable time discretization for the flow TL, we apply 
the recipe of the previous section with $F(L)=I+hL$, i.e. we take as 
a solution of this problem the map described by the discrete time Lax equation
\begin{equation}\label{dTL Lax}
\wL=\mbB^{-1}L\mbB=\mbA L\mbA^{-1} \quad{\rm with}\quad 
\mbB=\Pi_+(I+hL),\;\; \mbA=\Pi_-(I+hL)
\end{equation}

\begin{theorem}\label{discrete TL} {\rm\cite{S3} (see also \cite{GK2})}. 
Consider the change of variables $\cT(\ba,\bb)\mapsto\cT(a,b)$ defined by
the formulas
\begin{equation}\label{dTL loc map}
b_k=\bb_k+h\ba_{k-1},\qquad a_k=\ba_k(1+h\bb_k)
\end{equation}
The discrete time Lax equation {\rm(\ref{dTL Lax})} is equivalent to the map 
$(a,b)\mapsto(\wa,\wb)$ which in the variables $(\ba,\bb)$ is described by the 
following equations:
\begin{equation}\label{dTL loc}
\widetilde{\bb}_k=\bb_k+h(\ba_k-\widetilde{\ba}_{k-1}),\qquad 
\widetilde{\ba}_k(1+h\widetilde{\bb}_k)=\ba_k(1+h\bb_{k+1})
\end{equation}
\end{theorem}
\vspace{1.5mm}

\noindent
{\bf Proof.} The tri--diagonal structure of the matrix $L$ implies
the following bi--diagonal structure for the factors $\mbB, \mbA$:
\begin{equation}\label{dTL B}
\mbB=\Pi_+\Big(I+hL\Big)=
\sum_{k=1}^N(1+h\bb_k)E_{k,k}+h\lambda\sum_{k=1}^{N}E_{k+1,k}
\end{equation}
\begin{equation}\label{dTL A}
\mbA=\Pi_-\Big(I+hL\Big)=
I+h\lambda^{-1}\sum_{k=1}^{N}\ba_kE_{k,k+1}
\end{equation}
The formulas (\ref{dTL loc map}) represent the matrix equation $I+hT=\mbB\mbA$
which serves as a definition of the matrices $\mbB, \mbA$. Obviously, these 
formulas define a local diffeomorphism $\cT(\ba,\bb)\mapsto\cT(a,b)$. Now, the 
Lax equation (\ref{dTL Lax}) is
easy to see to be equivalent to $I+h\widetilde{T}=\mbB^{-1}(\mbB\mbA)\mbB=
\mbA(\mbB\mbA)\mbA^{-1}$, or
\begin{equation}\label{dTL Lax aux}
\widetilde{\mbB}\cdot\widetilde{\mbA}=\mbA\cdot\mbB
\end{equation}
The equations of motion (\ref{dTL loc}) are now nothing but the componentwise 
form of the latter matrix equation. \qed
\vspace{2mm}

The map (\ref{dTL loc}) will be denoted dTL. Obviously, it serves as a 
difference approximation to the Toda flow TL (\ref{TL}). 
The construction assures numerous positive properties of this discretization: 
in the coordinates $(a,b)$ the map dTL is Poisson with respect to each one of 
the Poisson brackets (\ref{TL l br}), (\ref{TL q br}), (\ref{TL c br}), it has 
the same integrals of motion as the flow TL, etc. Going to the coordinates
$(\ba,\bb)$ deforms the integrals of motion and the Poisson brackets. Moreover,
since the inverse to the map (\ref{dTL loc map}) is nonlocal, the invariant 
Poisson brackets become, generally speaking, also nonlocal in the coordinates 
$(\ba,\bb)$. Remarkably, there turn out to exist such linear combinations of 
the basic invariant Poisson brackets whose pull--backs to the coordinates 
$(\ba,\bb)$ are described by local formulas.

\begin{theorem}\label{dTL loc invariant PB}
{\rm a)} The pull--back of the bracket
\begin{equation}\label{dTL loc m1 br}
\{\cdot,\cdot\}_1+h\{\cdot,\cdot\}_2
\end{equation}
on $\cT(a,b)$ under the change of variables {\rm(\ref{dTL loc map})} is the
following bracket on $\cT(\ba,\bb)$:
\begin{equation}\label{dTL loc br 1}
\{\bb_k,\ba_k\}=-\ba_k(1+h\bb_k),\qquad \{\ba_k,\bb_{k+1}\}=-\ba_k(1+h\bb_{k+1})
\end{equation}
{\rm b)} The pull--back of the bracket 
\begin{equation}\label{dTL loc m2 br}
\{\cdot,\cdot\}_2+h\{\cdot,\cdot\}_3
\end{equation}
on $\cT(a,b)$ under the change of variables {\rm(\ref{dTL loc map})} is the
following bracket on $\cT(\ba,\bb)$:
\begin{eqnarray}
\{\bb_k,\ba_k\}=-\ba_k(\bb_k+h\ba_k)(1+h\bb_k),&\quad&
\{\ba_k,\bb_{k+1}\}=-\ba_k(\bb_{k+1}+h\ba_k)(1+h\bb_{k+1})\nonumber\\ 
\{\bb_k,\bb_{k+1}\}=-\ba_k(1+h\bb_k)(1+h\bb_{k+1}),&\quad&
\{\ba_k,\ba_{k+1}\}=-\ba_k\ba_{k+1}(1+h\bb_{k+1})\nonumber\\
\label{dTL loc br 2}
\end{eqnarray}
{\rm c)} The brackets {\rm(\ref{dTL loc br 1}), (\ref{dTL loc br 2})}
are compatible. The map {\rm dTL (\ref{dTL loc})} is Poisson with respect
to both of them.
\end{theorem}
{\bf Proof.} To prove the theorem, one has, for example, in the (less 
laborious) case a) to verify the following statement: the formulas 
(\ref{dTL loc br 1}) imply that the nonvanishing pairwise Poisson brackets 
of the functions (\ref{dTL loc map}) are 
\begin{eqnarray*}
\{b_k,a_k\}=-a_k(1+hb_k), & \qquad & \{a_k,b_{k+1}\}=-a_k(1+hb_{k+1})\\
\{b_k,b_{k+1}\}=-ha_k, &\qquad & \{a_k,a_{k+1}\}= -ha_ka_{k+1}
\end{eqnarray*}
This verification consists of straightforward calculations. \qed
\vspace{2mm}

The map (\ref{dTL loc}) was first found in \cite{HTI}, along with the Lax 
representation, but without  discussing its Poisson structure and its place 
in the continuous time Toda hierarchy.
\vspace{2mm}

We can now determine the hierarchy of continuous time lattice equations 
to which the map (\ref{dTL loc}) belongs. Clearly, this is the Toda hierarchy 
pulled--back under the map (\ref{dTL loc map}). The previous theorem allows 
to calculate the corresponding equations of motion in a systematic
(Hamiltonian) fashion.

\begin{theorem} {\rm \cite{K1}}. 
The pull--back of the flow {\rm TL} under the change of variables {\rm(\ref{dTL 
loc map})} is described by the following differential equations:
\begin{equation}\label{TL in loc map}
\dot{\bb}_k=(\ba_k-\ba_{k-1})(1+h\bb_k),\qquad 
\dot{\ba}_k=\ba_k(\bb_{k+1}-\bb_k)
\end{equation}
\end{theorem}
{\bf Proof.} To determine the pull--back of the flow TL, we can use the
Hamiltonian formalism. An opportunity to apply it is given by the Theorem
\ref{dTL loc invariant PB}. We shall use the statement a) only.
Consider the function $h^{-1}\rH_1(a,b)=h^{-1}\sum_{k=1}^N b_k$.
It is a Casimir of the bracket $\{\cdot,\cdot\}_1$, and generates exactly
the flow TL in the bracket $h\{\cdot,\cdot\}_2$. Hence it generates the flow
TL also in the bracket (\ref{dTL loc m1 br}). The pull--back of this Hamilton
function is equal to $h^{-1}\sum_{k=1}^N(\bb_k+h\ba_{k-1})$. It remains only 
to calculate the flow generated by this function in the Poisson brackets 
(\ref{dTL loc br 1}). This results in the equations of motion 
(\ref{TL in loc map}). \qed

\subsection{dTL $\to$ explicit dTL}
Now consider the equations (\ref{dTL loc}) as equations on the 2-dimensional
lattice. In other words, we attach the variables $\bb_k=\bb_k(t)$, 
$\ba_k=\ba_k(t)$ to the lattice site $(k,t)\in {\Bbb Z}\times h{\Bbb Z}$. 
A linear change of independent variables $(k,t)\mapsto (k,\tau)=(k,t+kh)$ 
mixes space and time coordinates, which is equivalent to changing 
the Cauchy path on the lattice to $\{\tau=0\}=\{t=-kh\}$. For another
instances of such staircase (or sawtooth) Cauchy paths on the lattice, see 
\cite{PNC}, \cite{NPCQ}, \cite{FV}. To deal with the new initial--value 
problem, denote
\begin{equation}\label{dTL to e}
\mbb_k(\tau)=\bb_k(t)=\bb_k(\tau-kh),\qquad \mba_k(\tau)=\ba_k(t)=\ba_k(\tau-kh)
\end{equation}
Denoting the $h$--shift in $\tau$ still by the tilde, it is easy to see that 
the variables $\mba_k=\mba_k(\tau)$, $\mbb_k=\mbb_k(\tau)$ satisfy the 
following difference equations:
\begin{equation}\label{dTL e}
\widetilde{\mbb}_k=\mbb_k+h(\mba_k-\mba_{k-1}),\qquad
\widetilde{\mba}_k(1+h\widetilde{\mbb}_k)=\mba_k(1+h\widetilde{\mbb}_{k+1})
\end{equation}
These equations serve as an {\it explicit} discretization of the flow TL. 
Indeed, they allow to calculate $(\widetilde{\mba},\widetilde{\mbb})$ 
explicitly, if $(\mba,\mbb)$ is known (first $\widetilde{\mbb}$, then
$\widetilde{\mba}$).

Of course, it is tempting to conclude about the integrability of the system
(\ref{dTL e}) from the known integrability of the equations (\ref{dTL loc}).
However, we do not know any general statements allowing such conclusions, so
that the integrability of (\ref{dTL e}) has to be proved independently. 
Moreover, we are not aware of any systematic procedure allowing
to determine invariant Poisson brackets of the map (\ref{dTL e}) 
starting with the known invariant Poisson brackets of the map
(\ref{dTL loc}). Hence even the {\it definition} of integrability is
nontrivial in the case of (\ref{dTL e}). Our approach to these problems will
be based on Lax representations.

Recall that the Lax equations may be considered as 
compatibility conditions of linear problems. In particular, the discrete time
Lax equation (\ref{dTL Lax aux}) has two versions:
\[
\widetilde{\mbB}\widetilde{\mbA}=\mbA(\mbB\mbA)\mbA^{-1}=\mbB^{-1}(\mbB\mbA)\mbB
\]
The first one is a compatibility condition of two linear problems:
\begin{equation}\label{dTL lin pr1 aux}
\mbB\mbA\psi=\mu\psi,\qquad \widetilde{\psi}=\mbA\psi
\end{equation}
while the second one is a compatibility condition of 
\begin{equation}\label{dTL lin pr2 aux}
\mbB\mbA\psi=\mu\psi,\qquad \widetilde{\psi}=\mbB^{-1}\psi
\end{equation}
Here $\psi=(\psi_1,\ldots,\psi_N)$ is an auxiliary vector, $\mu$ is a spectral
parameter (having nothing in common with the spectral parameter $\lambda$
entering the elements of the algebra $\g$ such as $L(a,b,\lambda)$).

These two pairs of linear problems are equivalent, correspondingly, to the
following ones:
\begin{equation}\label{dTL lin pr1}
\mbB\widetilde{\psi}=\mu\psi,\qquad \mbA\psi=\widetilde{\psi}
\end{equation}
and
\begin{equation}\label{dTL lin pr2}
\mbA\psi=\mu\widetilde{\psi},\qquad \mbB\widetilde{\psi}=\psi
\end{equation}
We are now in a position to derive from these two pairs of auxiliary linear
problems two different Lax representations for the map (\ref{dTL e}). To this
end, we assume that the variables $\psi_k(t)$ are also attached to lattice
sites $(k,t)$, and perform in the above equations the change of variables 
$(k,t)\mapsto(k,\tau)=(k,t+kh)$. We denote, in addition to the variables 
$\mba_k$, $\mbb_k$ introduced above, also
\[
\bpsi_k(\tau)=\psi_k(t)=\psi_k(\tau-kh)
\]

\begin{theorem} The map {\rm (\ref{dTL e})} allows the following Lax 
representation:
\begin{equation}\label{edTL Lax1}
\widetilde{\mbL}_-=\mbD^{-1}\mbL_-\mbD=\mbC^{-1}\mbL_-\mbC\;\Longleftrightarrow\;
\widetilde{\mbD}\,\widetilde{\mbC}^{-1}=\mbC^{-1}\mbD
\end{equation}
with the Lax matrix 
\begin{equation}\label{edTL Lax- matrix}
\mbL_-=(\mbD\mbC^{-1}-I)/h
=\left(\lambda\sum_{k=1}^NE_{k+1,k}+\sum_{k=1}^N(\mbb_k-h\mba_{k-1})E_{kk}
+\lambda^{-1}\sum_{k=1}^N\mba_kE_{k,k+1}\right)\mbC^{-1}
\end{equation}
where
\begin{eqnarray}
\mbD & = & \sum_{k=1}^N(1+h\mbb_k-h^2\mba_{k-1})E_{kk}+h\lambda
\sum_{k=1}^N E_{k+1,k}\\
\mbC & = & I-h\lambda^{-1}\sum_{k=1}^N\mba_kE_{k,k+1}
\end{eqnarray}
\end{theorem}
\vspace{1.5mm}

\noindent
{\bf Proof.} The statement may be, of course, easily {\it verified}, but we
want to show how it can be {\it derived}. To this end we start with
(\ref{dTL lin pr1}). In coordinates these equations read:
\[
(1+h\bb_k)\widetilde{\psi}_k+h\lambda\widetilde{\psi}_{k-1}=\mu\psi_k,\qquad
\widetilde{\psi}_k=\psi_k+h\lambda^{-1}\ba_k\psi_{k+1}
\]
This implies, after the change of variables $(k,t)\mapsto(k,\tau)$:
\[
(1+h\mbb_k)\widetilde{\bpsi}_k+h\lambda\bpsi_{k-1}=\mu\bpsi_k,\qquad
\widetilde{\bpsi}_k=\bpsi_k+h\lambda^{-1}\mba_k\widetilde{\bpsi}_{k+1}
\]
After simple manipulations the latter system may be put into the form
\[
(1+h\mbb_k-h^2\mba_{k-1})\widetilde{\bpsi}_k+h\lambda\widetilde{\bpsi}_{k-1}=
\mu\bpsi_k,\qquad \bpsi_k=\widetilde{\bpsi}_k-
h\lambda^{-1}\mba_k\widetilde{\bpsi}_{k+1}
\]
In matrix notations the latter equations look like
\[
\mbD\widetilde{\bpsi}=\mu\bpsi,\qquad \mbC\widetilde{\bpsi}=\bpsi
\]
with the matrices $\mbD$, $\mbC$ introduced above.
The compatibility condition of these two linear problems coincides with 
(\ref{edTL Lax1}). \qed
\vspace{1.5mm}

\begin{theorem} The map {\rm(\ref{dTL e})} allows the following Lax 
representation:
\begin{equation}\label{edTL Lax2}
\widetilde{\mbL}_+=\widetilde{\Delta}^{-1}\cF\cdot\mbL_+\cdot\cF^{-1}
\widetilde{\Delta}
\end{equation}
with the matrices
\begin{equation}
\mbL_+=\cF^{-1}\left(\lambda\sum_{k=1}^N E_{k+1,k}+
\sum_{k=1}^N(\mbb_k-h\mba_{k-1})E_{kk}+\lambda^{-1}\sum_{k=1}^N\mba_kE_{k,k+1}
\right)
\end{equation}
\begin{equation}\label{edTL F}
\cF=I-h\lambda\sum_{k=1}^N E_{k+1,k}=I-h\cE
\end{equation}
\begin{equation}
\Delta={\rm diag}(1+h\mbb_1,\ldots,1+h\mbb_N)
\end{equation}
\end{theorem}
{\bf Proof.} We perform with (\ref{dTL lin pr2}) transformations analogous to
those of the previous proof. In coordinates (\ref{dTL lin pr2}) reads:
\[
\psi_k+h\lambda^{-1}\ba_k\psi_{k+1}=\mu\widetilde{\psi}_k,\qquad
(1+h\bb_k)\widetilde{\psi}_k+h\lambda\widetilde{\psi}_{k-1}=\psi_k
\]
This implies, after the change of variables $(k,t)\mapsto(k,\tau)$:
\[
\bpsi_k+h\lambda^{-1}\mba_k\widetilde{\bpsi}_{k+1}=\mu\widetilde{\bpsi}_k,\qquad
(1+h\mbb_k)\widetilde{\bpsi}_k+h\lambda\bpsi_{k-1}=\bpsi_k
\]
Straightforward manipulations imply:
\[
\left(1+h\mbb_k-h^2\mba_k\,\frac{1+h\mbb_k}{1+h\mbb_{k+1}}\right)\bpsi_k+
h\lambda^{-1}\mba_k\,\frac{1+h\mbb_k}{1+h\mbb_{k+1}}\bpsi_{k+1}=\mu(\bpsi_k-
h\lambda\bpsi_{k-1}),
\]
\[
\widetilde{\bpsi}_k=\frac{1}{1+h\mbb_k}\,(\bpsi_k-h\lambda\bpsi_{k-1})
\]
This may be put in a matrix form:
\[
\cF^{-1}\mbP\bpsi=\mu\bpsi,\qquad \widetilde{\bpsi}=\Delta^{-1}\cF\bpsi
\]
with the matrices $\cF$, $\Delta$ as above, and
\begin{equation}
\mbP=\sum_{k=1}^N\left
(1+h\mbb_k-h^2\mba_k\,\frac{1+h\mbb_k}{1+h\mbb_{k+1}}\right)E_{kk}+
h\lambda^{-1}\sum_{k=1}^N\mba_k\,\frac{1+h\mbb_k}{1+h\mbb_{k+1}}\,E_{k,k+1}
\end{equation}
Now the compatibility condition of the latter two linear problems reads:
\[
\cF^{-1}\widetilde{\mbP}
=\Delta^{-1}\cF\cdot\cF^{-1}\mbP\cdot\cF^{-1}\Delta
\]
In order to bring this into the required form (\ref{edTL Lax2}), notice that 
the equations of motion (\ref{dTL e}) imply the following simple formula:
\begin{equation}
\widetilde{\mbP}=\sum_{k=1}^N(1+h\mbb_k-h^2\mba_{k-1})E_{kk}+
h\lambda^{-1}\sum_{k=1}^N\mba_kE_{k,k+1}
\end{equation}
so that
\[
\cF^{-1}\widetilde{\mbP}=I+h\mbL_+
\]
This completes the proof. \qed

\subsection{Explicit dTL $\to$ RTL}

Now that we have two different Lax representations for the map (\ref{dTL e}),
with the Lax matrices $\mbL_{\pm}$, the natural question is the one about 
their relation. It is easy to see that this relation may be described
as follows:
\[
\Omega_1\mbL_-(\lambda)\,\Omega_1^{-1}=
\mbL_+^{\rm T}(\alpha^{-1}\lambda^{-1})
\]
where $\alpha=(\mba_1\mba_2...\mba_N)^{1/N}$ and
\[
\Omega_1={\rm diag}(1,\alpha^{-1}\mba_1,\alpha^{-2}\mba_1\mba_2,\ldots,
\alpha^{-N+1}\mba_1\mba_2 ...\mba_{N-1})
\]
Both Lax matrices for the explicit dTL (\ref{dTL e}) are substantially 
different from the one for the implicit dTL (\ref{dTL loc}) (the latter Lax 
matrix is the usual Toda one (\ref{TL T}) with the variables $(a,b)$ 
parametrized according to (\ref{dTL loc map})). The next natural question is: 
what is the continuous time hierarchy, to which (\ref{dTL e}) belongs? In other 
words, how can one get continuous time flows with the Lax matrix $\mbL_{\pm}$,
and what is the underlying invariant Poisson structure(s)?

The answer is known for both Lax formulations. Introduce the phase space
\begin{equation}\label{RTL phase sp}
\cR={\Bbb R}^{2N}(\mbd_1,\mba_1,\ldots,\mbd_N,\mba_N)
\end{equation}
\begin{theorem} \label{RTL theorem-}
{\rm\cite{S2}} The Lax matrix map $\mbL_-(\mba,\mbd,\lambda):
\cR\mapsto\g$, where 
\begin{equation}\label{RTL Lax-}
\mbL_-=(\mbD\mbC^{-1}-I)/h=
\left(\lambda\sum_{k=1}^N E_{k+1,k}+\sum_{k=1}^N\mbd_kE_{kk}+\lambda^{-1}
\sum_{k=1}^N\mba_kE_{k,k+1}\right)\,\mbC^{-1}
\end{equation}
\begin{eqnarray}
\mbD(\mba,\mbd,\lambda) & = & \sum_{k=1}^N(1+h\mbd_k)E_{kk}
+h\lambda\sum_{k=1}^NE_{k+1,k}\\
\mbC(\mba,\mbd,\lambda) & = & I-h\lambda^{-1}\sum_{k=1}^N\mba_kE_{k,k+1}
\end{eqnarray}
is Poisson, if $\cR$ is equipped with the Poisson bracket
\begin{equation}\label{RTL l br}
\{\mbd_k,\mba_k\}_1=-\mba_k, \qquad \{\mba_k,\mbd_{k+1}\}_1=-\mba_k, \qquad
\{\mbd_k,\mbd_{k+1}\}_1=h\mba_k
\end{equation}
and $\g$ is equipped with ${\rm PB}_1(\rR)$, and also if $\cR$ is equipped 
with the Poisson bracket
\begin{equation}\label{RTL q br}
\begin{array}{cclcccl}
\{\mbd_k,\mba_k\}_2 & = & -\mba_k\mbd_k,&   \quad & 
\{\mba_k,\mbd_{k+1}\}_2 & = & -\mba_k\mbd_{k+1}  \\ \\
\{\mbd_k,\mbd_{k+1}\}_2 & = & -\mba_k,& \quad & 
\{\mba_k,\mba_{k+1}\}_2 & = & -\mba_k\mba_{k+1} 
\end{array}
\end{equation}
and $\g$ is equipped with ${\rm PB}_2(\rA_1,\rA_2,\rS)$. 
The hierarchy of continuous time flows is of the usual form
\begin{equation}\label{RTL hier}
\dot{\mbL}_-=\Big[\mbL_-,\pm\pi_{\pm}(f(\mbL_-))\Big]
\end{equation}
The ''first'' flow of the hierarchy, corresponding to $f(\mbL)=\mbL$,
coincides with {\rm RTL (\ref{RTL})} and allows a Lax representation with
the auxiliary matrix 
\begin{equation}\label{RTL Lax aux1}
\pi_+(\mbL_-)=\sum_{k=1}^N(\mbd_k+h\mba_{k-1})E_{kk}+
\lambda\sum_{k=1}^NE_{k+1,k}
\end{equation}
The map 
\[
\widetilde{\mbL}_-=\mbD^{-1}\mbL_-\mbD=\mbC^{-1}\mbL_-\mbC
\]
is interpolated by the flow of this hierarchy with $f(\mbL)=h^{-1}\log(I+h\mbL)$.
\end{theorem}
Actually, an additional information is available from \cite{S2}. The evolution
of the factors $\mbC$, $\mbD$ for the flows of the RTL hierarchy (\ref{RTL
hier}) is described by the Lax triads
\[
\dot{\mbD}=\pm\mbD\cdot\pi_{\pm}(f(\mbC^{-1}\mbD))
\mp\pi_{\pm}(f(\mbD\mbC^{-1}))\cdot\mbD
\]
\[
\dot{\mbC}=\pm\mbC\cdot\pi_{\pm}(f(\mbC^{-1}\mbD))
\mp\pi_{\pm}(f(\mbD\mbC^{-1}))\cdot\mbC
\] 
In particular, for the flow RTL (\ref{RTL}) the corresponding auxiliary 
matrices in the Lax triads are (\ref{RTL Lax aux1}) and
\begin{equation}\label{RTL Lax aux2}
\pi_+\Big((\mbC^{-1}\mbD-I)/h\Big)=\sum_{k=1}^N(\mbd_k+h\mba_k)E_{kk}+
\lambda\sum_{k=1}^NE_{k+1,k}
\end{equation}
For the case of the quadratic bracket an $r$--matrix interpretation in 
$\g\otimes\g$ is possible. Namely, the Lax matrix map $(\mbD,\mbC^{-1}):
\cR\mapsto\bg=\g\otimes\g$ is Poisson, if $\cR$ is equipped with the bracket 
$\{\cdot,\cdot\}_1+h\{\cdot,\cdot\}_2$, i.e.
\begin{equation}\label{RTL m br}
\{\mbd_k,\mba_k\}=-(1+h\mbd_k)\mba_k, \qquad 
\{\mba_k,\mbd_{k+1}\}=-(1+h\mbd_{k+1})\mba_k, \qquad
\{\mba_k,\mba_{k+1}\}=-h\mba_k\mba_{k+1}
\end{equation}
and $\g\otimes\g$ is equipped with 
$h{\rm PB}_2(\bA_1,\bA_2,\bS)$ with the operators
\begin{equation}\label{Ops in g+g}
\bA_1=\left(\begin{array}{cc} \rA_1 & -\rS\\ \rS^* & \rA_1\end{array}\right),
\qquad
\bA_2=\left(\begin{array}{cc} \rA_2 & -\rS^*\\ \rS & \rA_2\end{array}\right),
\qquad
\bS=\left(\begin{array}{cc} \rS & \rS\\ \rS & -\rS^* \end{array}\right)
\end{equation}
Theorem \ref{monodromy} then assures that the monodromy matrix maps
$\mbD\mbC^{-1}$ and $\mbC^{-1}\mbD$ are Poisson, if the target $\g$ is
equipped with $h{\rm PB}_2(\rA_1,\rA_2,\rS)$, which implies the Poisson
property for the Lax matrix maps $\mbL_-=(\mbD\mbC^{-1}-I)/h$ and
$(\mbC^{-1}\mbD-I)/h$, if the target $\g$ is equipped with ${\rm PB}_1(\rR)+
h{\rm PB}_2(\rA_1,\rA_2,\rS)$.
\vspace{2mm}

For the hierarchy attached to the Lax matrix $\mbL_+$, somewhat less detailed
information is available.
\begin{theorem} \label{RTL theorem+} {\rm\cite{BR}, \cite{GK1}} 
The hierarchy of Lax equations related to the Lax matrix 
\begin{equation}\label{RTL Lax+}
\mbL_+(\mba,\mbd,\lambda)=(I-h\cE)^{-1}
\left(\lambda\sum_{k=1}^N E_{k+1,k}+\sum_{k=1}^N\mbd_kE_{kk}+\lambda^{-1}
\sum_{k=1}^N\mba_kE_{k,k+1}\right)
\end{equation}
is given by the formula
\[
\dot{\mbL}_+=\Big[\mbL_+,\pm\pi_{\pm}(f(\mbL_+))+\sigma(f(\mbL_+))\Big]
\]
where $\sigma$ is a linear map from $\g$ into the set of diagonal matrices 
defined as
\[
\sigma\Big(\sum_p \ell^{(p)}\cE^p)=\sum_{p<0} h^{-p} \ell^{(p)}
\]
In particular, the ''first '' flow of this hierarchy corresponding to
$f(\mbL)=\mbL$ coincides with {\rm RTL (\ref{RTL})} and allows the Lax
representation 
\[
\dot{\mbL}_+=\Big[\cA,\mbL_+\Big]
\]
with the auxiliary matrix
\[
\cA=\pi_-(\mbL_+)-\sigma(\mbL_+)=\lambda^{-1}\sum_{k=1}^N\mba_kE_{k,k+1}-
h\sum_{k=1}^N \mba_kE_{kk}
\]
\end{theorem}
\vspace{2mm}

It has to be mentioned that each of the formulations above has its specific
advantages. Theorem \ref{RTL theorem-} has an advantage of including the 
RTL hierarchy into the standard framework of the lattice KP hierarchy. On the 
other hand, Theorem \ref{RTL theorem+}, found for RTL in \cite{BR}, 
has an advantage of being a particular case of a much more general construction, 
due to \cite{GK1}, which delivers a ''relativistic'' generalization 
of the whole lattice KP hierarchy. An $r$--matrix interpretation of the 
latter result is unknown (see, however, \cite{OR} for a linear bracket in the
open--end case).  

So, upon the identification
\[
\mbd_k=\mbb_k-h\mba_{k-1}
\]
the explicit discrete time Toda map (\ref{dTL e}) belongs to the relativistic
Toda hierarchy.
\vspace{2mm}

{\bf Remarks.} \begin{enumerate}
\item In the coordinates $(\mba,\mbb)$ the equations of motion (\ref{RTL})
take the form
\begin{equation}\label{RTL in ab}
\dot{\mbb}_k=\mba_k(1+h\mbb_k)-\mba_{k-1}(1+h\mbb_{k-1}), \qquad
\dot{\mba}_k=\mba_k(\mbb_{k+1}-\mbb_k+h\mba_{k+1}-h\mba_k)
\end{equation}
\item In the coordinates $(\mba,\mbb)$ the linear Poisson bracket 
(\ref{RTL l br}) of the RTL formally coincides (incidentally) with the linear 
Poisson bracket (\ref{TL l br}) of the usual Toda lattice:
\begin{equation}\label{RTL l br in ab}
\{\mbb_k,\mba_k\}_1=-\mba_k, \qquad \{\mba_k,\mbb_{k+1}\}_1=-\mba_k 
\end{equation}
Note also that in the coordinates $(\mba,\mbd)$ the quadratic Poisson
structure (\ref{RTL q br}) formally coincides with the quadratic Poisson
bracket (\ref{TL q br}) for the usual Toda lattice.
\item The third (''cubic'') invariant Poisson structure is known for the
RTL hierarchy, however its $r$--matrix interpretation is unknown:
\begin{equation}\label{RTL c br}
\begin{array}{ccl}
\{\mbd_k,\mba_k\}_3     & = & -\mba_k(\mbd_k^2+\mba_k+h\mbd\mba)\\ \\
\{\mba_k,\mbd_{k+1}\}_3 & = & -\mba_k(\mbd_{k+1}^2+\mba_k+h\mbd_{k+1}\mba)\\ \\ 
\{\mbd_k,\mbd_{k+1}\}_3 & = & -\mba_k(\mbd_k+\mbd_{k+1}+h\mbd_k\mbd_{k+1})\\ \\
\{\mba_k,\mba_{k+1}\}_3 & = & 
-\mba_k\mba_{k+1}(2\mbd_{k+1}+h\mba_k+h\mba_{k+1}) \\ \\
\{\mbd_k,\mba_{k+1}\}_3 & = & -\mba_k\mba_{k+1}(1+h\mbd_k)\\ \\
\{\mba_k,\mbd_{k+2}\}_3 & = & -\mba_k\mba_{k+1}(1+h\mbd_{k+2})\\ \\
\{\mba_k,\mba_{k+2}\}_3 & = & -h\mba_k\mba_{k+1}\mba_{k+2}
\end{array}
\end{equation}
\end{enumerate}

%%%%%%%%%%%%%%%%%%%%%%%%%%%%%%%%%%%%%%%%%%%%%%%%%%%%%%%%%%%%%%%%%%%%%%%%%

\setcounter{equation}{0}
\section{Volterra and relativistic Volterra lattices}
\subsection{VL}

We consider here the equations of motion of VL (\ref{VL}) under periodic 
boundary conditions, so that the corresponding phase space is:
\begin{equation}\label{VL phase sp}
\cV={\Bbb R}^{2N}(u_1,v_1,\ldots,u_N,v_N)
\end{equation}

There exist two compatible local Poisson brackets on $\cV$ invariant under 
the flow VL:
\begin{equation}\label{VL q br}
\{u_k,v_k\}_2=-u_kv_k,\qquad \{v_k,u_{k+1}\}_2=-v_ku_{k+1}
\end{equation}
and
\begin{equation}\label{VL c br}
\begin{array}{cclcccl}
\{u_k,v_k\}_3 & = & -u_kv_k(u_k+v_k), & \quad &
\{v_k,u_{k+1}\}_3 & = & -v_ku_{k+1}(v_k+u_{k+1})\\ \\
\{u_k,u_{k+1}\}_3 & = & -u_kv_ku_{k+1}, & \quad &
\{v_k,v_{k+1}\}_3 & = & -v_ku_{k+1}v_{k+1}
\end{array}
\end{equation}
The corresponding Hamilton functions for the flow VL are equal to
\begin{equation}\label{VL q Ham}
\rH_1(u,v)=\sum_{k=1}^N u_k+\sum_{k=1}^N v_k
\end{equation}
and
\begin{equation}\label{VL c Ham}
\rH_0(u,v)=\sum_{k=1}^N \log(u_k)\qquad{\rm or}\qquad
\rH_0(u,v)=\sum_{k=1}^N \log(v_k)
\end{equation}
(the difference of the latter two functions is a Casimir of $\{\cdot,\cdot\}_3$).

The Lax representation of VL we use here lives in $\bg=\g\otimes\g$ and was
introduced in \cite{K1} (where the system ({\ref{VL}) was called ''modified
Toda lattice''), see also \cite{S5}. Consider the following two matrices:
\begin{equation}\label{VL U}
U(u,v,\lambda)=\lambda\sum_{k=1}^N E_{k+1,k}+\sum_{k=1}^N u_kE_{k,k}=\cE+u
\end{equation}
\begin{equation}\label{VL V} 
V(u,v,\lambda)=I+\lambda^{-1}\sum_{k=1}^N v_kE_{k,k+1}=I+v\cE^{-1}
\end{equation}
These formulas define the ''Lax matrix'' $(U,V):\cV\mapsto\g\otimes\g$. 
The flow {\rm(\ref{VL})} is equivalent to either of the following Lax 
equations in  $\g\otimes\g$:
\begin{equation}\label{VL Lax}
\left\{\begin{array}{l}\dot{U}=UB_2-B_1U \\ \dot{V}=VB_1-B_2V \end{array}\right.
\qquad{\rm or}\qquad
\left\{\begin{array}{l}\dot{U} =A_1U-UA_2 \\ \dot{V}=A_2V-VA_1 \end{array}\right.
\end{equation}
with the matrices
\begin{equation}
B_1=\pi_+(UV),\qquad B_2=\pi_+(VU),\qquad A_1=\pi_-(UV),\qquad
A_2=\pi_-(VU)
\end{equation}
so that
\begin{eqnarray}
B_1 & = & \lambda\sum_{k=1}^NE_{k+1,k}+\sum_{k=1}^N (u_k+v_{k-1})E_{kk}
\label{VL B1} \\
B_2 & = & \lambda\sum_{k=1}^NE_{k+1,k}+\sum_{k=1}^N (u_k+v_k)E_{kk}
\label{VL B2} \\
A_1 & = & \lambda^{-1}\sum_{k=1}^N u_kv_kE_{k,k+1} \label{VL A1} \\
A_2 & = & \lambda^{-1}\sum_{k=1}^N u_{k+1}v_kE_{k,k+1} \label{VL A2}
\end{eqnarray}
As a consequence, the matrices 
\begin{equation}\label{VL T}
T_1(u,v,\lambda)=U(u,v,\lambda)V(u,v,\lambda),\qquad
T_2(u,v,\lambda)=V(u,v,\lambda)U(u,v,\lambda)
\end{equation}
satisfy the usual Lax equations in $\g$:
\begin{equation}\label{VL Lax in g}
\dot{T}=[T,\pm\pi_{\pm}(T)]
\end{equation}
The Lax equations (\ref{VL Lax}), (\ref{VL Lax in g}) may be given an 
$r$--matrix interpretation. 
\begin{theorem} {\rm\cite{S5}} The Lax matrix map $(U,V):\cV\mapsto\g\otimes\g$
is Poisson, if $\cV$ is equipped with the bracket $\{\cdot,\cdot\}_2$ and
$\g\otimes\g$ is equipped with the bracket ${\rm PB}_2(\bA_1,\bA_2,\bS)$ 
corresponding to the operators $\bA_1,\bA_2,\bS$ defined in {\rm(\ref{Ops 
in g+g})}. The maps $T_{1,2}:\g\otimes\g\mapsto\g$, 
\[
T_1(U,V)=UV, \qquad T_2(U,V)=VU
\]
are Poisson, if the target space $\g$ is equipped with the bracket
${\rm PB}_2(\rA_1,\rA_2,\rS)$
\end{theorem}
It is easy to see that the matrices $T_1,T_2$ formally coincide with the
Lax matrix (\ref{TL T}) of the Toda lattice, with the coordinates $(a,b)$
given by the corresponding formula for the Miura maps in (\ref{Miuras}).
The Poisson property of the monodromy map is therefore reflected in the
(first half) of the following statements about the Miura maps.
\begin{theorem} The Miura maps $\cM_{1,2}:\cV\mapsto\cT$ are Poisson,
if $\cV$ carries the bracket $\{\cdot,\cdot\}_2$ {\rm(\ref{VL q br})} and
$\cT$ carries the bracket $\{\cdot,\cdot\}_2$ {\rm(\ref{TL q br})}, and also
if $\cV$ carries the bracket $\{\cdot,\cdot\}_3$ {\rm(\ref{VL c br})} and
$\cT$ carries the bracket $\{\cdot,\cdot\}_3$ {\rm(\ref{TL c br})}.
\end{theorem}

\subsection{VL $\to$ dVL}

To find an integrable time discretization for the flow VL, we 
apply the recipe of Sect. \ref{Sect recipe} with $F(T)=I+hT$, 
i.e. we consider the map described by the
discrete time ''Lax triads''
\begin{equation}\label{dVL Lax in g+g}
\left\{\begin{array}{l}\wU=\mbB_1^{-1}U\mbB_2 \\ \wV=\mbB_2^{-1}V\mbB_1
\end{array}\right. \qquad{\rm or}\qquad
\left\{\begin{array}{l}\wU=\mbA_1U\mbA_2^{-1} \\ \wV=\mbA_2V\mbA_1^{-1}
\end{array}\right.
\end{equation}
with
\[
\mbB_1=\Pi_+(I+hUV), \quad \mbB_2=\Pi_+(I+hVU), \quad \mbA_1=\Pi_-(I+hUV),
\quad \mbA_2=\Pi_-(I+hVU)
\]
\begin{theorem}\label{discrete VL}
Consider the change of variables $\cV(\bu,\bv)\mapsto\cV(u,v)$ defined by
the formulas
\begin{equation}\label{dVL loc map}
u_k=\bu_k(1+h\bv_{k-1}), \qquad v_k=\bv_k(1+h\bu_k)
\end{equation}
The discrete time Lax equations  {\rm(\ref{dVL Lax in g+g})}
are equivalent to the map $(u,v)\mapsto(\wu,\wv)$ which in coordinates
$(\bu,\bv)$ is described by the following equations:
\begin{equation}\label{dVL loc}
\widetilde{\bu}_k(1+h\widetilde{\bv}_{k-1})=\bu_k(1+h\bv_k),\qquad
\widetilde{\bv}_k(1+h\widetilde{\bu}_k)=\bv_k(1+h\bu_{k+1})
\end{equation}
\end{theorem}
{\bf Proof.} The formulas (\ref{dVL loc map}) allow to find the factors
$\mbB_{1,2}$, $\mbA_{1,2}$ in a closed form. Indeed, with the help of 
(\ref{dVL loc map}) we can represent (\ref{VL U}), (\ref{VL V}) as
\begin{eqnarray} 
U & = & \sum_{k=1}^N\bu_k(1+h\bv_{k-1})E_{kk}+\lambda\sum_{k=1}^N
E_{k+1,k} \label{dVL U}\\
V & = & I+\lambda^{-1}\sum_{k=1}^N \bv_k(1+h\bu_k)E_{k,k+1}
\label{dVL V}
\end{eqnarray}
From these formulas we derive:
\begin{eqnarray*}
I+hUV & = & h\lambda\sum_{k=1}^NE_{k+1,k}+
\sum_{k=1}^N\Big((1+h\bu_k)(1+h\bv_{k-1})+h^2\bu_{k-1}\bv_{k-1}\Big)E_{kk}\\
& & +h\lambda^{-1}\sum_{k=1}^N \bu_k\bv_k(1+h\bu_k)(1+h\bv_{k-1})E_{k,k+1}
\end{eqnarray*}
Obviously, this matrix may be factorized as $I+hUV=\mbB_1\mbA_1$, where
\begin{eqnarray}
\mbB_1 & = & 
\sum_{k=1}^N(1+h\bu_k)(1+h\bv_{k-1})E_{kk}+h\lambda\sum_{k=1}^{N}E_{k+1,k}
\label{dVL B1}\\
\mbA_1 & = & 
I+h\lambda^{-1}\sum_{k=1}^{N}\bu_k\bv_kE_{k,k+1}
\label{dVL A1}
\end{eqnarray}
Similarly, we find:
\begin{eqnarray*}
I+hVU  & = & h\lambda\sum_{k=1}^NE_{k+1,k}+\sum_{k=1}^N\Big((1+h\bu_k)(1+h\bv_k)
+h^2\bu_k\bv_{k-1}\Big)E_{kk}\\
& & +h\lambda^{-1}\sum_{k=1}^N \bu_{k+1}\bv_k(1+h\bu_k)(1+h\bv_k)E_{k,k+1}
\end{eqnarray*}
which may be factorized as $I+hVU=\mbB_2\mbA_2$, where
\begin{eqnarray}
\mbB_2 & = & 
\sum_{k=1}^N(1+h\bu_k)(1+h\bv_k)E_{kk}+h\lambda\sum_{k=1}^{N}E_{k+1,k}
\label{dVL B2}\\
\mbA_2 & = & 
I+h\lambda^{-1}\sum_{k=1}^{N}\bu_{k+1}\bv_kE_{k,k+1}
\label{dVL A2}
\end{eqnarray}
Now the discrete time Lax equations $\mbB_1\wU=U\mbB_2$, $\mbB_2\wV=V\mbB_1$ 
(or their $\mbA$--analogs) immediately imply the equations of motion 
(\ref{dVL loc}). \qed
\vspace{2mm}

Comparing two pairs of formulas (\ref{dVL B1}), (\ref{dVL A1}) and
(\ref{dVL B2}), (\ref{dVL A2}) for dVL with the formulas (\ref{dTL B}),
(\ref{dTL A}), we see immediately that the Miura maps $\cM_{1,2}:
\cV(u,v)\mapsto\cT(a,b)$ are conjugated by the changes of variables
(\ref{dVL loc map}), (\ref{dTL loc map}) with the maps
$\bM_{1,2}:\cV(\bu,\bv)\mapsto\cT(\ba,\bb)$ given by:
\begin{equation}\label{Miuras loc}
\bM_1:\left\{\begin{array}{l} 1+h\bb_k=(1+h\bu_k)(1+h\bv_{k-1})\\ 
\ba_k=\bu_k\bv_k\end{array}\right.
\qquad
\bM_2:\left\{\begin{array}{l} 1+h\bb_k=(1+h\bu_k)(1+h\bv_k)\\ 
\ba_k=\bu_{k+1}\bv_k\end{array}\right.
\end{equation}

The map (\ref{dVL loc}), denoted hereafter by dVL, serves as a difference 
approximation to the flow VL (\ref{VL}). By construction, in the variables
$(u,v)$ this map shares the integrals of motion and the invariant Poisson 
structures with the flow VL. In the coordinates $(\bu,\bv)$ the integrals
of motion and the Poisson brackets become deformed, the latter ones become
nonlocal, since the inverse of the map (\ref{dVL loc map}) is nonlocal. 
Nevertheless, we have the following local invariant Poisson bracket for the 
map dVL:

\begin{theorem}\label{local PB for local dVL}  {\rm\cite{K1}}
The pull--back of the bracket
\begin{equation}\label{dVL loc br}
\{\cdot,\cdot\}_2+h\{\cdot,\cdot\}_3
\end{equation} 
on $\cV(u,v)$ under the change of variables {\rm(\ref{dVL loc map})} is 
the following bracket on $\cV(\bu,\bv)$:
\begin{equation}\label{dVL loc PB}
\{\bu_k,\bv_k\}=-\bu_k\bv_k(1+h\bu_k)(1+h\bv_k),\qquad
\{\bv_k,\bu_{k+1}\}=-\bv_k\bu_{k+1}(1+h\bv_k)(1+h\bu_{k+1})
\end{equation}
The map {\rm(\ref{dVL loc})} is Poisson with respect to the bracket 
{\rm (\ref{dVL loc PB})}. The maps $\bM_{1,2}$ {\rm(\ref{Miuras loc})}
are Poisson, if $\cV(\bu,\bv)$ carries the bracket {\rm(\ref{dVL loc PB})}
and $\cT(\ba,\bb)$ carries the bracket {\rm(\ref{dTL loc br 2})}.
\end{theorem}
{\bf Proof} -- by a direct verification. \qed
\vspace{2mm}

The map (\ref{dVL loc}) was found in \cite{THO}, \cite{HTI}, where, however,
its place in the continuous time Volterra hierarchy and its Poisson structure
were not elaborated. The Miura maps (\ref{Miuras loc}) were found in \cite{K1}
without connecting them to the discretization problem, and also in \cite{HT}
in the discretization context, but without Poisson properties.
\vspace{1.5mm}

The previous theorem allows also to calculate the equations of motion of
continuous time hierarchy, to which the map dVL belongs.

\begin{theorem} \label{VL in local dVL coords} {\rm\cite{K1}}
The pull--back of the flow {\rm VL} under the map {\rm(\ref{dVL loc map})} is
described by the following equations of motion:
\begin{equation}\label{VL in loc dVL map}
\dot{\bu}_k=\bu_k(1+h\bu_k)(\bv_k-\bv_{k-1}),\qquad
\dot{\bv}_k=\bv_k(1+h\bv_k)(\bu_{k+1}-\bv_k)
\end{equation}
\end{theorem}
{\bf Proof.}  Since the flow VL has a Hamilton function
$h^{-1}\,\log\rH_0(u,v)$ in the Poisson bracket $h\{\cdot,\cdot\}_3$, and
this function is a Casimir of the bracket $\{\cdot,\cdot\}_2$, we conclude
that this function generates the flow VL also in the bracket (\ref{dVL loc br}).
This means that the pull--back of the flow VL is a Hamiltonian flow  in the
bracket (\ref{dVL loc PB}) with the Hamilton function
\[
h^{-1}\sum_{k=1}^N\log\Big(\bu_k(1+h\bv_{k-1})\Big)
\]
Calculating the corresponding equations of motion, we arrive at the equations
(\ref{VL in loc dVL map}). \qed

\subsection{dVL $\to$ explicit dVL}
We use the same trick as in the case of dTL, in order to extract from dVL
the explicit discretization for VL. Namely, we attach the variables $\bu_k=
\bu_k(t)$, $\bv_k=\bv_k(t)$ to the lattice site $(k,t)\in{\Bbb Z}\times 
h{\Bbb Z}$, and then perform the change of independent variables $(k,t)\mapsto 
(k,\tau)=(k,t+kh)$. We denote
\begin{equation}
\mbu_k(\tau)=\bu_k(t)=\bu_k(\tau-kh), \qquad
\mbv_k(\tau)=\bv_k(t)=\bv_k(\tau-kh)
\end{equation}
Denoting the $h$--shift in $\tau$ still by the tilde, we immediately derive
from (\ref{dVL loc}) the following difference equations for the variables
$\mbu_k$, $\mbv_k$:
\begin{equation}\label{dVL e}
\widetilde{\mbu}_k(1+h\mbv_{k-1})=\mbu_k(1+h\mbv_k),\qquad
\widetilde{\mbv}_k(1+h\widetilde{\mbu}_k)=\mbv_k(1+h\widetilde{\mbu}_{k+1})
\end{equation}
This is clearly an {\it explicit} discretization, since, knowing $(\mbu,\mbv)$,
it allows to calculate explicitly first $\widetilde{\mbu}$, and then
$\widetilde{\mbv}$.

We find now Lax representations for the map (\ref{dVL e}). As in the case 
of the dTL, there exist two versions thereof.
\begin{theorem}\label{Lax representation- edVL} 
The map {\rm(\ref{dVL e})} allows the following Lax 
representation:
\begin{equation}\label{edVL Lax- triad}
\widetilde{\mbU}=\mbC^{-1}\mbU\mbC_2,\qquad 
\widetilde{\mbV}\,\widetilde{\mbC}^{-1}=\mbC_2^{-1}\mbV
\end{equation}
with the matrices
\begin{eqnarray}
\mbU(\mbu,\mbv,\lambda) & = & 
\sum_{k=1}^N\mbu_kE_{kk}+\lambda\sum_{k=1}^NE_{k+1,k} \label{edVL U}\\
\mbV(\mbu,\mbv,\lambda) & = & I+\lambda^{-1}\sum_{k=1}^N\mbv_kE_{k,k+1}
\label{edVL V}\\
\mbC(\mbu,\mbv,\lambda) & = & I-h\lambda^{-1}\sum_{k=1}^N\mbu_k\mbv_kE_{k,k+1}
\label{edVL C} \\
\mbC_2(\mbu,\mbv,\lambda) & = & I-h\lambda^{-1}\sum_{k=1}^N
\widetilde{\mbu}_{k+1}\mbv_kE_{k,k+1} \label{edVL C2}
\end{eqnarray}
The Lax representation {\rm (\ref{edVL Lax- triad})} implies also
\begin{equation}\label{edVL Lax-}
\widetilde{\mbL}_-=\mbC^{-1}\mbL_-\mbC\;\Longleftrightarrow\;
\widetilde{\mbU}\,\widetilde{\mbV}\,\widetilde{\mbC}^{-1}=\mbC^{-1}\mbU\mbV
\end{equation}
with the Lax matrix 
\begin{equation}\label{edVL Lax matrix-}
\mbL_-=\mbU\mbV\mbC^{-1}
\end{equation} 
\end{theorem}
{\bf Proof.} We start with the following Lax representation of the dVL in
$\g\otimes\g$:
\[
\widetilde{U}=\mbA_1U\mbA_2^{-1},\qquad \widetilde{V}=\mbA_2V\mbA_1^{-1}
\]
with the matrices (\ref{dVL U}), (\ref{dVL V}), (\ref{dVL A1}), (\ref{dVL A2}).
These Lax equations may be considered as the compatibility condition of the 
following linear problems:
\[
\left\{\begin{array}{l} U\phi=\mu\psi \\ \\ V\psi=\mu\phi \end{array}\right. 
\qquad
\left\{\begin{array}{l} \widetilde{\phi}=\mbA_2\phi \\ \\ \widetilde{\psi}=
\mbA_1\psi\end{array}\right.
\]
We shall call the first pair {\it the spectral linear problem}, and the
second pair {\it the evolutionary linear problem}. In coordinates the above
equations may be presented as
\[
\left\{\begin{array}{l}
\bu_k(1+h\bv_{k-1})\phi_k+\lambda\phi_{k-1}=\mu\psi_k\\ \\
\psi_k+\lambda^{-1}\bv_k(1+h\bu_k)\psi_{k+1}=\mu\phi_k
\end{array}\right.\qquad
\left\{\begin{array}{l}
\widetilde{\phi}_k=\phi_k+h\lambda^{-1}\bu_{k+1}\bv_k\phi_{k+1}\\ \\
\widetilde{\psi}_k=\psi_k+h\lambda^{-1}\bu_k\bv_k\phi_{k+1}
\end{array}\right.
\]
Upon the change of variables $(k,t)\mapsto(k,\tau)$ this implies:
\[
\left\{\begin{array}{l}
\mbu_k(1+h\undertilde{\mbv_{k-1}})\bphi_k+\lambda\undertilde{\bphi_{k-1}}
=\mu\bpsi_k\\ \\
\bpsi_k+\lambda^{-1}\mbv_k(1+h\mbu_k)\widetilde{\bpsi}_{k+1}=\mu\bphi_k
\end{array}\right.\qquad
\left\{\begin{array}{l}
\widetilde{\bphi}_k=\bphi_k+h\lambda^{-1}\widetilde{\mbu}_{k+1}\mbv_k
\widetilde{\bphi}_{k+1}\\ \\
\widetilde{\bpsi}_k=\bpsi_k+h\lambda^{-1}\mbu_k\mbv_k\widetilde{\bphi}_{k+1}
\end{array}\right.
\]
After simple manipulations (substitute into the first pair of equalities
the values of $\undertilde{\bphi_{k-1}}$ and $\bpsi_k$ obtained from the 
second pair of equalities) this can be brought into the form
\[
\left\{\begin{array}{l}
\mbu_k\bphi_k+\lambda\bphi_{k-1}=\mu\bpsi_k\\ \\
\widetilde{\bpsi}_k+\lambda^{-1}\mbv_k\widetilde{\bpsi}_{k+1}=\mu\bphi_k
\end{array}\right.\qquad
\left\{\begin{array}{l}
\bphi_k=\widetilde{\bphi}_k-h\lambda^{-1}\widetilde{\mbu}_{k+1}\mbv_k
\widetilde{\bphi}_{k+1}\\ \\
\bpsi_k=\widetilde{\bpsi}_k-h\lambda^{-1}\mbu_k\mbv_k\widetilde{\bphi}_{k+1}
\end{array}\right.
\]
or in the matrix form
\[
\left\{\begin{array}{l}
\mbU\bphi=\mu\bpsi\\ \\ \mbV\widetilde{\bpsi}=\mu\bphi\end{array}\right.\qquad
\left\{\begin{array}{l}
\mbC_2\widetilde{\bphi}=\bphi\\ \\
\mbC\widetilde{\bpsi}=\bpsi
\end{array}\right.
\]
where we introduced the matrices $\mbU$, $\mbV$, $\mbC$, $\mbC_2$ from 
(\ref{edVL U})--(\ref{edVL C2}). The compatibility condition of these linear 
problems coincides with (\ref{edVL Lax- triad}). \qed
\vspace{2mm}

\begin{theorem} \label{Lax representation+ edVL}
The map {\rm(\ref{dVL e})} allows the following Lax
representation:
\begin{equation}\label{edVL Lax+ triad}
\left\{
\begin{array}{rcl}
\cF^{-1}\widetilde{\mbU}\widetilde{\Delta}_2
&=&(\widetilde{\Delta}_1^{-1}\Delta_3^{-1}\cF)\cdot
\cF^{-1}\mbU\Delta_2\cdot(\cF^{-1}\widetilde{\Delta}_1\widetilde{\Delta}_2)\\ \\
\widetilde{\Delta}_2^{-1}\widetilde{\mbV}&=&
(\widetilde{\Delta}_2^{-1}\widetilde{\Delta}_1^{-1}\cF)\cdot
\Delta_2^{-1}\mbV\cdot(\cF^{-1}\Delta_3\widetilde{\Delta}_1)
\end{array}\right.
\end{equation}
with the matrices $\mbU$, $\mbV$, $\cF$ from {\rm (\ref{edVL U}), (\ref{edVL V}),
(\ref{edTL F})}, and
\begin{equation}\label{edVL Del 1}
\Delta_1={\rm diag}(1+h\mbu_k), \qquad \Delta_2={\rm diag}(1+h\mbv_k)
\end{equation}
\begin{equation}\label{edVL Del 3}
\Delta_3 ={\rm diag}(1+h\mbv_{k-1})  
\end{equation} 
The Lax representation {\rm(\ref{edVL Lax+ triad})} implies also
\begin{equation}\label{edVL Lax+}
\widetilde{\mbL}_+=(\widetilde{\Delta}_1^{-1}\Delta_3^{-1}\cF)\cdot\mbL_+\cdot
(\cF^{-1}\Delta_3\widetilde{\Delta}_1)
\end{equation} 
with the Lax matrix
\begin{equation}\label{edVL Lax matrix+}
\mbL_+=\cF^{-1}\mbU\mbV
\end{equation} 
\end{theorem}
{\bf Proof.} This time we start with the following Lax representation
of the map dVL:
\[
\widetilde{U}=\mbB_1^{-1}U\mbB_2, \qquad \widetilde{V}=\mbB_2^{-1}V\mbB_1
\]
with the matrices (\ref{dVL U}), (\ref{dVL V}), (\ref{dVL B1}), (\ref{dVL B2}).
These Lax equations are the compatibility conditions of the following linear
problems:
\[
\left\{\begin{array}{l} \widetilde{U}\widetilde{\phi}=\mu\widetilde{\psi}\\ \\
\widetilde{V}\widetilde{\psi}=\mu\widetilde{\phi}\end{array}\right.\qquad
\left\{\begin{array}{l} \mbB_2\widetilde{\phi}=\phi\\ \\
\mbB_1\widetilde{\psi}=\psi\end{array}\right.
\]
(it turns out to be more convenient to take the tilded versions of the 
equations in the first pair). In components the above equations take
the form:
\[
\left\{\begin{array}{l}
\widetilde{\bu}_k(1+h\widetilde{\bv}_{k-1})\widetilde{\phi}_k+\lambda
\widetilde{\phi}_{k-1}=\mu\widetilde{\psi}_k \\ \\
\widetilde{\psi}_k+\lambda^{-1}\widetilde{\bv}_k(1+h\widetilde{\bu}_k)
\widetilde{\psi}_{k+1}=\mu\widetilde{\phi}_k\end{array}\right.\qquad
\left\{\begin{array}{l}
(1+h\bu_k)(1+h\bv_k)\widetilde{\phi}_k+h\lambda\widetilde{\phi}_{k-1}=\phi_k\\
\\ 
(1+h\bv_{k-1})(1+h\bu_k)\widetilde{\psi}_k+h\lambda\widetilde{\psi}_{k-1}
=\psi_k
\end{array}\right.
\]
After the change of variables $(k,t)\mapsto(k,\tau)$ this takes the form:
\[
\left\{\begin{array}{l}
\widetilde{\mbu}_k(1+h\mbv_{k-1})\widetilde{\bphi}_k+\lambda
\bphi_{k-1}=\mu\widetilde{\bpsi}_k \\ \\
\widetilde{\bpsi}_k+\lambda^{-1}\widetilde{\mbv}_k(1+h\widetilde{\mbu}_k)
\widetilde{\widetilde{\bpsi}}_{k+1}=\mu\widetilde{\bphi}_k\end{array}\right.
\qquad \left\{\begin{array}{l}
(1+h\mbu_k)(1+h\mbv_k)\widetilde{\bphi}_k+h\lambda\bphi_{k-1}=\bphi_k\\
\\ 
(1+h\undertilde{\mbv_{k-1}})(1+h\mbu_k)\widetilde{\bpsi}_k+
h\lambda\bpsi_{k-1}=\bpsi_k
\end{array}\right.
\]
Now we use the equations of motion (\ref{dVL e}) to bring the first pair
(the spectral problem) into the form
\begin{equation}\label{edVL lax+ aux}
\left\{\begin{array}{l}
\mbu_k(1+h\mbv_k)\widetilde{\bphi}_k+\lambda
\bphi_{k-1}=\mu\widetilde{\bpsi}_k \\ \\
\widetilde{\bpsi}_k+\lambda^{-1}\mbv_k(1+h\widetilde{\mbu}_{k+1})
\widetilde{\widetilde{\bpsi}}_{k+1}=\mu\widetilde{\bphi}_k\end{array}\right.
\end{equation}
The second pair (the evolutionary problem) may be presented as
\begin{equation}\label{edVL lax+ aux1}
\left\{\begin{array}{l}
(1+h\mbu_k)(1+h\mbv_k)\widetilde{\bphi}_k=\bphi_k-h\lambda\bphi_{k-1}\\
\\ 
(1+h\undertilde{\mbv_{k-1}})(1+h\mbu_k)\widetilde{\bpsi}_k=\bpsi_k
-h\lambda\bpsi_{k-1}
\end{array}\right.\quad\Longleftrightarrow\quad
\left\{\begin{array}{l}
\widetilde{\bphi}=\Delta_2^{-1}\Delta_1^{-1}\cF\bphi \\ \\
\widetilde{\bpsi}=\Delta_1^{-1}\undertilde{\!\Delta_3^{-1}}\cF\bpsi
\end{array}\right.
\end{equation}
Now we transform the equations describing the spectral problems. 
From the first equations of the two pairs in (\ref{edVL lax+ aux}),
(\ref{edVL lax+ aux1}) it is easy to derive:
\[
\bphi_k-(1+h\mbv_k)\widetilde{\bphi}_k=h\mu\widetilde{\bpsi}_k
\]
and using this in the first equation in (\ref{edVL lax+ aux}), we bring the
latter into the form
\begin{equation}\label{edVL lax+ aux2}
\mbu_k(1+h\mbv_k)\widetilde{\bphi}_k
+h\lambda(1+h\mbv_{k-1})\widetilde{\bphi}_{k-1}=
\mu(\widetilde{\bpsi}_k-h\lambda\widetilde{\bpsi}_{k-1})
\end{equation}
From the second equation in (\ref{edVL lax+ aux1}) we find
the value 
\[
(1+h\mbu_{k+1})\widetilde{\widetilde{\bpsi}}_{k+1}=
\frac{1}{1+h\mbv_k}(\widetilde{\bpsi}_{k+1}-h\lambda\widetilde{\bpsi}_k)
\]
which, being substituted into the second spectral equation in 
(\ref{edVL lax+ aux}), implies:
\begin{equation}\label{edVL lax+ aux3}
\frac{1}{1+h\mbv_k}\Big(\widetilde{\bpsi}_k+
\lambda^{-1}\mbv_k\widetilde{\bpsi}_{k+1}\Big)=\mu\widetilde{\bphi}_k
\end{equation}
Now we can put the spectral problems in (\ref{edVL lax+ aux2}), 
(\ref{edVL lax+ aux3}) into the matrix form:
\begin{equation}\label{edVL lax+ aux4}
\left\{\begin{array}{l}
\cF^{-1}\mbU\Delta_2\widetilde{\bphi}=\mu\widetilde{\bpsi}\\ \\
\Delta_2^{-1}\mbV\widetilde{\bpsi}=\mu\widetilde{\bphi}
\end{array}\right.\qquad
\end{equation}
The compatibilty conditions of the linear problems (\ref{edVL lax+ aux1}),
(\ref{edVL lax+ aux4}) coincide with (\ref{edVL Lax+}). \qed

\subsection{Explicit dVL $\to$ RVL}

Now we have obtained two Lax matrices for the explicit dVL:
\[
\mbL_-(\mbu,\mbv,\lambda)=\mbU\mbV\mbC^{-1}\quad{\rm and}\quad
\mbL_+(\mbu,\mbv,\lambda)=\cF^{-1}\mbU\mbV
\]
The relation between them is the following:
\[
\Omega_1\mbU(\lambda)\,\Omega_2^{-1}=\mbV^{\rm T}(\lambda'),
\qquad 
\Omega_2\mbV(\lambda)\,\Omega_1^{-1}=\mbU^{\rm T}(\lambda')
\qquad
\Omega_1\mbC(\lambda)\,\Omega_1^{-1}=\cF^{\rm T}(\lambda')
\]
so that 
\[
\Omega_1\mbL_-(\lambda)\,\Omega_1^{-1}=\mbL_+^{\rm T}(\lambda')
\]
where $\lambda'=\alpha^{-1}\lambda^{-1}$, 
$\alpha=(\mbu_1\mbv_1...\mbu_N\mbv_N)^{1/N}$ and
\begin{eqnarray*}
\Omega_1 & = & {\rm diag}(1,\alpha^{-1}\mbu_1\mbv_1,
\alpha^{-2}\mbu_1\mbv_1\mbu_2\mbv_2,\ldots,
\alpha^{-N+1}\mbu_1\mbv_1...\mbu_{N-1}\mbv_{N-1})\\
\Omega_2 & = & {\rm diag}(1,\alpha^{-1}\mbv_1\mbu_2,
\alpha^{-2}\mbv_1\mbu_2\mbv_2\mbu_3,\ldots,
\alpha^{-N+1}\mbv_1\mbu_2...\mbv_{N-1}\mbu_N)
\end{eqnarray*}
We {\it postulate} that the hypothetic relativistic Volterra lattice (RVL) 
lives in the hierarchy associated to these Lax matrices, and it remains to 
identify this hierarchy. (The above relation between $\mbL_{\pm}$ assures
that the both Lax matrices generate one and the same hierarchy). We start with 
the Lax matrix $\mbL_-$.
\begin{theorem} The triples $(\mbU,\mbC^{-1},\mbV)$ form a Poisson Lax matrix
map $\cV(\mbu,\mbv)\mapsto\g\otimes\g\otimes\g$, if $\cV(\mbu,\mbv)$ carries
the Poisson bracket
\begin{equation}\label{RVL q br}
\{\mbu_k,\mbv_k\}_2=-\mbu_k\mbv_k,\qquad
\{\mbv_k,\mbu_{k+1}\}_2=-\mbv_k\mbu_{k+1}
\end{equation}
and $\g\otimes\g\otimes\g$ carries the Poisson bracket ${\rm PB}_2(\bA_1,
\bA_2,\bS)$ defined by the operators
\begin{equation}\label{Ops in g+g+g}
\bA_1=\left(\begin{array}{ccc} \rA_1 & -\rS & -\rS\\ 
                               \rS^* & \rA_1 & \rS^*\\
                               \rS^* & -\rS & \rA_1
\end{array}\right),
\qquad
\bA_2=\left(\begin{array}{ccc} \rA_2 & -\rS^* & -\rS^*\\ 
                               \rS & \rA_2 & -\rS^* \\
                               \rS & \rS & \rA_2
\end{array}\right),
\qquad
\bS=\left(\begin{array}{ccc} \rS & \rS & \rS\\ 
                             \rS & -\rS^* & -\rS^*\\
                             \rS & \rS & -\rS^*
\end{array}\right)
\end{equation}
The Lax matrix maps $\mbL_-=\mbU\mbV\mbC^{-1}$, $\mbV\mbC^{-1}\mbU$,
$\mbC^{-1}\mbU\mbV:\cV(\mbu,\mbv)\mapsto\g$ are Poisson, if the target $\g$
is equipped with ${\rm PB}_2(\rA_1,\rA_2,\rS)$. The hierarchy of continuous
time flows is described by the Lax triads
\begin{eqnarray}
\dot{\mbU} & = & \pm\mbU\cdot\pi_{\pm}\Big(f(\mbV\mbC^{-1}\mbU)\Big)
\mp\pi_{\pm}\Big(f(\mbU\mbV\mbC^{-1})\Big)\cdot\mbU\\
\dot{\mbC} & = & \pm\mbC\cdot\pi_{\pm}\Big(f(\mbC^{-1}\mbU\mbV)\Big)
\mp\pi_{\pm}\Big(f(\mbU\mbV\mbC^{-1})\Big)\cdot\mbC\\
\dot{\mbV} & = & \pm\mbV\cdot\pi_{\pm}\Big(f(\mbC^{-1}\mbU\mbV)\Big)
\mp\pi_{\pm}\Big(f(\mbV\mbC^{-1}\mbU)\Big)\cdot\mbV
\end{eqnarray}
As a consequence, the evolution of the matrix $\mbL_-=\mbU\mbV\mbC^{-1}$ is
described by the standard Lax equation
\[
\dot{\mbL}_-=\Big[\mbL_-,\pm\pi_{\pm}(f(\mbL_-))\Big]
\]
In particular, the ''first'' flow corresponding to $f(\mbL)=\mbL$, coincides 
with the {\rm RVL (\ref{RVL+})}, and the auxiliary matrices in the Lax triads
are given by
\begin{eqnarray}
\pi_+(\mbU\mbV\mbC^{-1}) & = & \sum_{k=1}^N(\mbu_k+\mbv_{k-1}+
h\mbu_{k-1}\mbv_{k-1})E_{kk}+\lambda\sum_{k=1}^NE_{k+1,k}\\
\pi_+(\mbC^{-1}\mbU\mbV) & = & \sum_{k=1}^N(\mbu_k+\mbv_{k-1}+
h\mbu_k\mbv_k)E_{kk}+\lambda\sum_{k=1}^NE_{k+1,k}\\
\pi_+(\mbV\mbC^{-1}\mbU) & = & \sum_{k=1}^N(\mbu_k+\mbv_k+
h\mbu_k\mbv_k)E_{kk}+\lambda\sum_{k=1}^NE_{k+1,k}
\end{eqnarray}
The map {\rm (\ref{dVL e})} belongs to this hierarchy (in particular, is
Poisson with respect to the bracket {\rm(\ref{RVL q br})}) and corresponds to
$f(\mbL)=h^{-1}\log(I+h\mbL)$.
\end{theorem}
{\bf Proof.} The first statement is proved with the help of the tensor 
notations for the quadratic $r$--matrix brackets, along the same lines
as the proof of analogous statements in \cite{S2}, \cite{S5}. All other
statements, except the last one, are consequences of the first one and 
Theorem \ref{monodromy}. To identify the place of the map (\ref{dVL e})
in this hierarchy, we have, referring to Theorem \ref{Lax representation- edVL},
to prove the following relations:
\[
\mbC^{-1}=\Pi_-(I+h\mbU\mbV\mbC^{-1}),\quad
\mbC_2^{-1}=\Pi_-(I+h\mbV\mbC^{-1}\mbU),\quad
\widetilde{\mbC}^{-1}=\Pi_-(I+h\mbC^{-1}\mbU\mbV)
\]
Clearly, it is enough to prove only the first of them, but it follows
immediately from
\[
I+h\mbU\mbV\mbC^{-1}=(\mbC+h\mbU\mbV)\mbC^{-1}
\]
and
\[
\mbC+h\mbU\mbV=\sum_{k=1}^N(1+h\mbu_k+h\mbv_{k-1})E_{kk}+h\lambda
\sum_{k=1}^NE_{k+1,k}\in\G_+,\qquad \mbC^{-1}\in\G_-
\]
The proof is complete. \qed
\vspace{2mm}

Turning to the case of the Lax matrix $\mbL_+$, we have the following results.
\begin{theorem} \label{RVL hierarchy+}
The hierarchy associated with the Lax matrix 
\[
\mbL_+=\cF^{-1}\mbU\mbV
\]
is described by the following Lax triads:
\begin{equation}\label{RVL Lax triads+}
\dot{\mbU}=\cA_1\mbU-\mbU\cA_2,\qquad \dot{\mbV}=\cA_2\mbV-\mbV\cA_0
\end{equation}
which have to be supplemented by the identity
\begin{equation}\label{RVL Lax triads+ add}
\dot{\cF}=0=\cA_1\cF-\cF\cA_0
\end{equation}
so that for the matrix $\mbL_+=\cF^{-1}\mbU\mbV$ there holds the Lax equation
\begin{equation}\label{RVL+ Lax+}
\dot{\mbL}_+=\Big[\cA_0,\mbL_+\Big]
\end{equation}
Here 
\begin{equation}\label{RVL+ A}
\cA_0=\pi_-(f(\mbL_+))-\sigma(f(\mbL_+))
\end{equation}
and the matrices $\cA_{1,2}$ are uniquely defined by the condition of 
compatibility of the equations of the Lax triads {\rm (\ref{RVL Lax triads+}),
(\ref{RVL Lax triads+ add})}. In particular, the ''first'' flow of the
hierarchy corresponding to $f(\mbL)=\mbL$, coincides with {\rm RVL (\ref{RVL+})},
and for this flow
\begin{eqnarray}
\cA_0 & = & \lambda^{-1}\sum_{k=1}^N\mbu_k\mbv_kE_{k,k+1}-
h\sum_{k=1}^N\mbu_k\mbv_kE_{kk} \\
\cA_1 & = & \lambda^{-1}\sum_{k=1}^N\mbu_k\mbv_kE_{k,k+1}-
h\sum_{k=1}^N\mbu_{k-1}\mbv_{k-1}E_{kk}\\
\cA_2 & = & \lambda^{-1}\sum_{k=1}^N\mbu_{k+1}\mbv_kE_{k,k+1}-
h\sum_{k=1}^N\mbu_k\mbv_kE_{kk}
\end{eqnarray}
\end{theorem}
{\bf Proof} goes by a direct check. We define $\cA_0$ by (\ref{RVL+ A}), then
put $\cA_1=\cF\cA_0\cF^{-1}\in\g_0\oplus\g_-$. The matrix 
$\cA_2\in\g_0\oplus\g_-$ 
may be defined in two ways to make either of two Lax triads in (\ref{RVL Lax 
triads+}) consistent. These two definitions have to be consistent, because
the choice (\ref{RVL+ A}) assures the validity of the Lax equation (\ref{RVL+ 
Lax+}) for the matrix $\mbL_+$, due to Theorem \ref{RTL theorem+}. \qed 
\vspace{2mm}

Comparing the Lax matrices $\mbL_{\pm}$ for the RTL hierarchy (see
(\ref{RTL Lax-}, (\ref{RTL Lax+})) and for the RVL hierarchy (see
(\ref{edVL Lax-}), (\ref{edVL Lax+})), we see that the Miura map between
two hierarchies formally coincides with the Miura map between the
TL hierarchy and the VL hierarchy, because it is equivalent to identifying
$L(\mba,\mbd,\lambda)$ with $U(\mbu,\mbv,\lambda)V(\mbu,\mbv,\lambda)$,
where $L,U,V$ are the matrices from the nonrelativistic theories. This Miura
map is
\[
\cM_1(\mbu,\mbv)=(\mba,\mbd):\;\left\{\begin{array}{l}
\mbd_k=\mbu_k+\mbv_{k-1}\\\mba_k=\mbu_k\mbv_k\end{array}\right.
\]
This Miura map is Poisson with respect to the invariant quadratic Poisson
brackets, since these also formally coincide with the invariant quadratic
Poisson brackets of the nonrelativistic hierarchies. Recall that also the
cubic Poisson bracket (\ref{RTL c br}) is known for the RTL hierarchy.
It turns out that its pull--back by the Miura map is given by the following
nice formulas:
\begin{eqnarray}\label{RVL c br}
\{\mbu_k,\mbv_k\}_3=-\mbu_k\mbv_k(\mbu_k+\mbv_k+h\mbu_k\mbv_k),&\quad&
\{\mbv_k,\mbu_{k+1}\}_3=-\mbv_k\mbu_{k+1}(\mbv_k+\mbu_{k+1})
\nonumber\\
\{\mbu_k,\mbu_{k+1}\}_3=-\mbu_k\mbv_k\mbu_{k+1}(1+h\mbu_k),&\quad&
\{\mbv_k,\mbv_{k+1}\}_3=-\mbv_k\mbu_{k+1}\mbv_{k+1}(1+h\mbv_{k+1})
\nonumber\\
\{\mbv_k,\mbu_{k+2}\}_3=-h\mbv_k\mbu_{k+1}\mbv_{k+1}\mbu_{k+2}&&
\end{eqnarray}

Finally, let us comment on the twin RVL flow (\ref{RVL-}). Obviously, 
this system, as well as the corresponding Miura map $\cM_2$, may be obtained 
upon the renaming $\mbu_k\to\mbv_k$, $\mbv_k\to\mbu_{k+1}$. The Lax matrices 
for the corresponding hierarchy arise from the ones previously concerned
upon change $\mbU\mbV$ to $\mbV\mbU$. The explicit discretization of
the VL living in this hierarchy reads:
\begin{equation}
\widetilde{\mbu}_k(1+h\widetilde{\mbv}_{k-1})=\mbu_k(1+h\widetilde{\mbv}_k),
\qquad \widetilde{\mbv}_k(1+h\mbu_k)=\mbv_k(1+h\mbu_{k+1})
\end{equation}
(here, knowing $\mbu$, $\mbv$, one calculates first $\widetilde{\mbv}$, 
and then $\widetilde{\mbu}$).

%%%%%%%%%%%%%%%%%%%%%%%%%%%%%%%%%%%%%%%%%%%%%%%%%%%%%%%%%%%%%%%%%%%%%%%%
\setcounter{equation}{0}
\section{Bogoyavlensky lattices}
\subsection{BL2}

A class of systems called Bogoyavlensky lattices (although discovered several 
times independently before his papers \cite{B}, see the references in \cite{S5},
and also \cite{K1}) serve as generalizations of the Volterra lattice.
We consider here only one representative of this class, which is however
typical.

The integrable lattice system considered in the present section
will be called BL2. Its phase space (in the periodic case) is
\begin{equation}\label{BL2 phase sp}
\cB={\Bbb R}^{3N}(u_1,v_1,w_1,\ldots,u_N,v_N,w_N)
\end{equation}
The equations of motion read:
\begin{eqnarray} 
\dot{u}_k & = & u_k(v_k+w_k-v_{k-1}-w_{k-1}) \nonumber\\
\dot{v}_k & = & v_k(u_{k+1}+w_k-u_k-w_{k-1}) \label{BL2}\\
\dot{w}_k & = & w_k(u_{k+1}+v_{k+1}-u_k-v_k) \nonumber
\end{eqnarray}
This system is Hamiltonian with the local Poisson brackets on $\cB$
\begin{eqnarray}
\{u_k,v_k\}_2=-u_kv_k,& \quad & \{v_k,u_{k+1}\}_2=-v_ku_{k+1} \nonumber\\
\{u_k,w_k\}_2=-u_kw_k,& \quad & \{w_k,u_{k+1}\}_2=-w_ku_{k+1} \label{BL2 q br}\\
\{v_k,w_k\}_2=-v_kw_k,& \quad & \{w_k,v_{k+1}\}_2=-w_kv_{k+1} \nonumber
\end{eqnarray}
and the Hamilton function
\begin{equation}\label{BL2 q Ham}
\rH_1(u,v,w)=\sum_{k=1}^N u_k+\sum_{k=1}^N v_k+\sum_{k=1}^N w_k
\end{equation}
Notice that the system BL2 is more familiar in the form
\[
\dot{a}_k=a_k(a_{k+1}+a_{k+2}-a_{k-1}-a_{k-2})
\]
which arises upon the re-naming of variables
\[
u_k=a_{3k-2},\qquad v_k=a_{3k-1},\qquad w_k=a_{3k}
\]

We use here the Lax representation of BL2 living in $\bg=\g\otimes\g\otimes\g$ 
and introduced in \cite{S5}. Consider the following three matrices:
\begin{eqnarray}
U(u,v,w,\lambda) & = &
\lambda\sum_{k=1}^N E_{k+1,k}+\sum_{k=1}^N u_kE_{k,k}=\cE+u \label{BL2 U}\\
V(u,v,w,\lambda) & = & I+\lambda^{-1}\sum_{k=1}^N v_kE_{k,k+1}=I+v\cE^{-1}
\label{BL2 V} \\
W(u,v,w,\lambda) & = & I+\lambda^{-1}\sum_{k=1}^N w_kE_{k,k+1}=I+w\cE^{-1}
\label{BL2 W}
\end{eqnarray}
These formulas define the ''Lax matrix'' $(U,V,W):\cB\mapsto
\g\otimes\g\otimes\g$. The flow {\rm(\ref{BL2})} is equivalent to either of 
the following Lax equations in  $\g\otimes\g\otimes\g$:
\begin{equation}\label{BL2 Lax}
\left\{\begin{array}{l}\dot{U}=UB_3-B_1U \\ \dot{V}=VB_1-B_2V \\
\dot{W}=WB_2-B_3W \end{array}\right.
\qquad{\rm or}\qquad
\left\{\begin{array}{l}\dot{U} =A_1U-UA_3 \\ \dot{V}=A_2V-VA_1 \\
\dot{W}=A_3V-VA_2\end{array}\right.
\end{equation}
with the matrices
\begin{equation}
B_1=\pi_+(UWV),\qquad B_2=\pi_+(VUW),\qquad B_3=\pi_+(WVU)
\end{equation}
\begin{equation}
A_1=\pi_-(UWV),\qquad A_2=\pi_-(VUW),\qquad A_3=\pi_-(WVU)
\end{equation}
so that
\begin{eqnarray}
B_1 & = & \lambda\sum_{k=1}^NE_{k+1,k}+
\sum_{k=1}^N (u_k+v_{k-1}+w_{k-1})E_{kk}\\
B_2 & = & \lambda\sum_{k=1}^NE_{k+1,k}+
\sum_{k=1}^N (u_k+v_k+w_{k-1})E_{kk}\\
B_3 & = & \lambda\sum_{k=1}^NE_{k+1,k}+
\sum_{k=1}^N (u_k+v_k+w_k)E_{kk}\\
A_1 & = & \lambda^{-1}\sum_{k=1}^N (w_{k-1}v_k+u_kv_k+u_kw_k)
E_{k,k+1}+\lambda^{-2}\sum_{k=1}^N u_kw_kv_{k+1}E_{k,k+2}\nonumber\\
A_2 & = & \lambda^{-1}\sum_{k=1}^N (u_kw_k+v_kw_k+v_ku_{k+1})
E_{k,k+1}+\lambda^{-2}\sum_{k=1}^N v_ku_{k+1}w_{k+1}E_{k,k+2}\nonumber\\
A_3 & = & \lambda^{-1}\sum_{k=1}^N (v_ku_{k+1}+w_ku_{k+1}+
w_kv_{k+1})E_{k,k+1}+\lambda^{-2}\sum_{k=1}^N w_kv_{k+1}u_{k+2}E_{k,k+2}
\nonumber
\end{eqnarray}
The Lax equations (\ref{BL2 Lax}) allow the following $r$--matrix 
interpretation. 
\begin{theorem} {\rm\cite{S5}} The Lax matrix map $(U,V,W):\cB\mapsto
\g\otimes\g\otimes\g$ is Poisson, if $\cB$ is equipped with the bracket 
$\{\cdot,\cdot\}_2$ and $\g\otimes\g\otimes\g$ is equipped with the bracket 
${\rm PB}_2(\bA_1,\bA_2,\bS)$ corresponding to the operators $\bA_1,\bA_2,\bS$ 
defined in {\rm(\ref{Ops in g+g+g})}. The monodromy maps 
$T_{1,2,3}:\g\otimes\g\otimes\g\mapsto\g$, 
\[
T_1(U,V,W)=UWV, \qquad T_2(U,V,W)=VUW, \qquad T_3(U,V,W)=WVU
\] 
are Poisson, if the target space $\g$ is equipped with the bracket 
${\rm PB}_2(\rA_1,\rA_2,\rS)$.
\end{theorem}

\subsection{BL2 $\to$ dBL2}

To find an integrable time discretization for the flow BL2, we again
apply the recipe of Sect. \ref{Sect recipe} with $F(T)=I+hT$, 
i.e. we consider the map described by the
discrete time ''Lax triads''
\begin{equation}\label{dBL2 Lax in g+g+g}
\left\{\begin{array}{l}\wU=\mbB_1^{-1}U\mbB_3 \\ \wV=\mbB_2^{-1}V\mbB_1 \\
\wW=\mbB_3^{-1}W\mbB_2 \end{array}\right. \qquad{\rm or}\qquad
\left\{\begin{array}{l}\wU=\mbA_1U\mbA_3^{-1} \\ \wV=\mbA_2V\mbA_1^{-1} \\
\wW=\mbA_3W\mbA_2^{-1}
\end{array}\right.
\end{equation}
with
\[
\mbB_1=\Pi_+(I+hUWV), \quad \mbB_2=\Pi_+(I+hVUW), \quad \mbB_3=\Pi_+(I+hWVU)
\]
\[
\mbA_1=\Pi_-(I+hUWV), \quad \mbA_2=\Pi_-(I+hVUW), \quad \mbA_3=\Pi_-(I+hWVU)
\]
\begin{theorem}\label{discrete BL2}
Consider the change of variables $\cB(\bu,\bv,\bw)\mapsto\cB(u,v,w)$ defined by
the formulas
\begin{eqnarray}\label{dBL2 loc map}
u_k & = & \bu_k(1+h\bw_{k-1})(1+h\bv_{k-1}) \nonumber\\
v_k & = & \bv_k(1+h\bu_k)(1+h\bw_{k-1})\\
w_k & = & \bw_k(1+h\bv_k)(1+h\bu_k) \nonumber
\end{eqnarray}
The discrete time Lax equations  {\rm(\ref{dBL2 Lax in g+g+g})}
are equivalent to the map $(u,v,w)\mapsto(\wu,\wv,\ww)$ which in coordinates
$(\bu,\bv,\bw)$ is described by the following equations:
\begin{eqnarray}\label{dBL2 loc}
\widetilde{\bu}_k(1+h\widetilde{\bv}_{k-1})(1+h\widetilde{\bw}_{k-1}) & = & 
\bu_k(1+h\bv_k)(1+h\bw_k) \nonumber\\
\widetilde{\bv}_k(1+h\widetilde{\bu}_k)(1+h\widetilde{\bw}_{k-1}) & = &
\bv_k(1+h\bu_{k+1})(1+h\bw_k) \\
\widetilde{\bw}_k(1+h\widetilde{\bu}_k)(1+h\widetilde{\bv}_k) & = &
\bw_k(1+h\bu_{k+1})(1+h\bv_{k+1}) \nonumber
\end{eqnarray}
\end{theorem}
{\bf Proof} is parallel to the proof of Theorem \ref{discrete VL}.
The formulas (\ref{dBL2 loc map}) allow to find the factors
$\mbB_{1,2,3}$, $\mbA_{1,2,3}$ in a closed form. Indeed, with the help of 
(\ref{dBL2 loc map}) we can represent (\ref{BL2 U})--(\ref{BL2 W}) as
\begin{eqnarray} 
U & = & \sum_{k=1}^N\bu_k(1+h\bw_{k-1})(1+h\bv_{k-1})E_{kk}+
\lambda\sum_{k=1}^N E_{k+1,k} \label{dBL2 U}\\
V & = & I+\lambda^{-1}\sum_{k=1}^N \bv_k(1+h\bu_k)(1+h\bw_{k-1})
E_{k,k+1} \label{dBL2 V}\\
W & = & I+\lambda^{-1}\sum_{k=1}^N \bw_k(1+h\bv_k)(1+h\bu_k)
E_{k,k+1} \label{dBL2 W}
\end{eqnarray}
From these formulas we derive the following expressions for the factors
$\mbB_i=\Pi_+(I+hT_i)$ and $\mbA_i=\Pi_-(I+hT_i)$, $i=1,2,3$:
\begin{eqnarray}
\mbB_1 & = & 
\sum_{k=1}^N (1+h\bu_k)(1+h\bw_{k-1})(1+h\bv_{k-1})E_{kk}
+h\lambda\sum_{k=1}^{N}E_{k+1,k}  \label{dBL2 B1}\\
\mbB_2 & = & 
\sum_{k=1}^N (1+h\bv_k)(1+h\bu_k)(1+h\bw_{k-1})E_{kk}+
h\lambda\sum_{k=1}^{N}E_{k+1,k}  \label{dBL2 B2}\\
\mbB_3 & = & 
\sum_{k=1}^N (1+h\bw_k)(1+h\bv_k)(1+h\bu_k)E_{kk}+
h\lambda\sum_{k=1}^{N}E_{k+1,k}  \label{dBL2 B3}\\
\mbA_1 & = & 
I+h\lambda^{-1}\sum_{k=1}^{N}(\bw_{k-1}\bv_k+\bu_k\bv_k+\bu_k\bw_k+
h\bw_{k-1}\bu_k\bv_k+h\bu_k\bv_k\bw_k)E_{k,k+1}\nonumber\\
 & & +h\lambda^{-2}\sum_{k=1}^N \bu_k\bw_k\bv_{k+1}(1+h\bv_k)
(1+h\bw_k)(1+h\bu_{k+1})E_{k,k+2}  \label{dBL2 A1}\\
\mbA_2 & = & 
I+h\lambda^{-1}\sum_{k=1}^{N}(\bu_k\bw_k+\bv_k\bw_k+\bv_k\bu_{k+1}+
h\bu_k\bv_k\bw_k+h\bv_k\bw_k\bu_{k+1})E_{k,k+1}\nonumber\\
 & & +h\lambda^{-2}\sum_{k=1}^N \bv_k\bu_{k+1}\bw_{k+1}(1+h\bw_k)
(1+h\bu_{k+1})(1+h\bv_{k+1})E_{k,k+2}  \label{dBL2 A2}\\
\mbA_3 & = & 
I+h\lambda^{-1}\sum_{k=1}^{N}(\bv_k\bu_{k+1}+\bw_k\bu_{k+1}+\bw_k\bv_{k+1}+
h\bv_k\bw_k\bu_{k+1}+h\bw_k\bu_{k+1}\bv_{k+1})E_{k,k+1}\nonumber\\
 & & +h\lambda^{-2}\sum_{k=1}^N \bw_k\bv_{k+1}\bu_{k+2}(1+h\bu_{k+1})
(1+h\bv_{k+1})(1+h\bw_{k+1})E_{k,k+2}
\label{dBL2 A3}
\end{eqnarray}
Now the equations of motion (\ref{dBL2 loc}) follow directly from the 
$\mbB$--version of the discrete time Lax triads (\ref{dBL2 Lax in g+g+g}). 
\qed
\vspace{2mm}

The map (\ref{dBL2 loc}), denoted hereafter by dBL2, serves as a difference 
approximation to the flow BL2 (\ref{BL2}). According to the general
properties of discretizations of Sect. \ref{Sect recipe}, this map
in the coordinates $(u,v,w)$ is Poisson with respect to the bracket
(\ref{BL2 q br}). Unfortunately, this bracket becomes highly nonlocal
in the coordinates $(\bu,\bv,\bw)$. After the re-naming
\[
\bu_k=\ba_{3k-2},\qquad \bv_k=\ba_{3k-1},\qquad \bw_k=\ba_{3k}
\]
the map dBL2 turns into
\[
\widetilde{\ba}_k(1+h\widetilde{\ba}_{k-1})(1+h\widetilde{\ba}_{k-2})
=\ba_k(1+h\ba_{k+1})(1+h\ba_{k+2})
\]
and in this form it appeared for the first time in \cite{THO}, and then
in \cite{PN}, \cite{S4}. 

\subsection{dBL2 $\to$ explicit dBL2}
Following the scheme used already for dTL and dVL, we extract an explicit
discretization for the flow BL2 from the map dBL2. To this end we denote
\begin{eqnarray}
\mbu_k(\tau) & = & \bu_k(t)=\bu_k(\tau-kh) \nonumber\\
\mbv_k(\tau) & = & \bv_k(t)=\bv_k(\tau-kh) \\
\mbw_k(\tau) & = & \bw_k(t)=\bw_k(\tau-kh) \nonumber
\end{eqnarray}
These variables satisfy the following difference equations with respect to
the discrete time $\tau$:
\begin{eqnarray}\label{dBL2 e}
\widetilde{\mbu}_k(1+h\mbv_{k-1})(1+h\mbw_{k-1}) & = & 
\mbu_k(1+h\mbv_k)(1+h\mbw_k) \nonumber\\
\widetilde{\mbv}_k(1+h\widetilde{\mbu}_k)(1+h\mbw_{k-1}) & = &
\mbv_k(1+h\widetilde{\mbu}_{k+1})(1+h\mbw_k) \\
\widetilde{\mbw}_k(1+h\widetilde{\mbu}_k)(1+h\widetilde{\mbv}_k) & = &
\mbw_k(1+h\widetilde{\mbu}_{k+1})(1+h\widetilde{\mbv}_{k+1})
\end{eqnarray}
This is indeed an explicit discretization, because, knowing $(\mbu,\mbv,\mbw)$,
one calculates explicitly $\widetilde{\mbu}$, $\widetilde{\mbv}$,
$\widetilde{\mbw}$ (in this order). We now turn to finding a Lax representation
for this map. Unlike the previous cases, we could find reasonable formulas
only starting from the $\mbB$--version of the Lax representation for dBL2.
\begin{theorem} \label{Lax representation+ edBL2}
The map {\rm(\ref{dBL2 e})} allows the following Lax
representation:
\begin{equation}\label{edBL2 Lax+ triad}
\left\{
\begin{array}{rcl}
\cF^{-1}\widetilde{\mbU}\widetilde{\Delta}_2\widetilde{\Delta}_3 & = & 
(\widetilde{\Delta}_1^{-1}\Delta_4^{-1}\Delta_5^{-1}\cF)\cdot
\cF^{-1}\mbU\Delta_2\Delta_3\cdot
(\cF^{-1}\widetilde{\Delta}_1\widetilde{\Delta}_2\widetilde{\Delta}_3)\\ \\
\widetilde{\Delta}_2^{-1}\widetilde{\mbV} & = & 
(\widetilde{\Delta}_2^{-1}\widetilde{\Delta}_1^{-1}\Delta_5^{-1}\cF)
\cdot\Delta_2^{-1}\mbV\cdot(\cF^{-1}\Delta_5\Delta_4\widetilde{\Delta}_1) \\ \\
\widetilde{\Delta}_3^{-1}\widetilde{\mbW} & = & 
(\widetilde{\Delta}_3^{-1}\widetilde{\Delta}_2^{-1}\widetilde{\Delta}_1^{-1}\cF)
\cdot\Delta_3^{-1}\mbW\cdot(\cF^{-1}\Delta_5\widetilde{\Delta}_1
\widetilde{\Delta}_2)
\end{array}\right.
\end{equation}
with the matrices 
\begin{eqnarray}
\mbU & = & \lambda\sum_{k=1}^N E_{k+1,k}+\sum_{k=1}^N \mbu_kE_{k,k}
\label{edBL2 U}\\
\mbV & = & I+\lambda^{-1}\sum_{k=1}^N \mbv_kE_{k,k+1} \label{edBL2 V} \\
\mbW & = & I+\lambda^{-1}\sum_{k=1}^N \mbw_kE_{k,k+1} \label{edBL2 W}
\end{eqnarray}
and
\begin{equation}\label{edBL2 Del 123}
\Delta_1={\rm diag}(1+h\mbu_k), \qquad
\Delta_2={\rm diag}(1+h\mbv_k), \qquad
\Delta_3={\rm diag}(1+h\mbw_k)
\end{equation}
\begin{equation}\label{edBL2 Del 45}
\Delta_4={\rm diag}(1+h\mbv_{k-1}), \qquad
\Delta_5={\rm diag}(1+h\mbw_{k-1}) 
\end{equation} 
The Lax representation {\rm(\ref{edBL2 Lax+ triad})} implies also
\begin{equation}\label{edBL2 Lax+}
\widetilde{\mbL}_+=(\widetilde{\Delta}_1^{-1}\Delta_4^{-1}\Delta_5^{-1}\cF)
\cdot \mbL_+ \cdot (\cF^{-1}\Delta_5\Delta_4\widetilde{\Delta}_1)
\end{equation} 
with the Lax matrix
\begin{equation}\label{edBL2 Lax matrix+}
\mbL_+=(\cF^{-1}\mbU\Delta_2\Delta_3)(\Delta_3^{-1}\mbW)(\Delta_2^{-1}\mbV)
=\cF^{-1}\mbU\Delta_2\mbW\Delta_2^{-1}\mbV
\end{equation} 
\end{theorem}
{\bf Proof.} We start with the following Lax representation of the map dBL2:
\[
\widetilde{U}=\mbB_1^{-1}U\mbB_3, \qquad \widetilde{V}=\mbB_2^{-1}V\mbB_1,
\qquad \widetilde{W}=\mbB_3^{-1}W\mbB_2
\]
with the matrices (\ref{dBL2 U})--(\ref{dBL2 W}), 
(\ref{dBL2 B1})--(\ref{dBL2 B3}).
These Lax equations are the compatibility conditions of the following linear
problems:
\[
\left\{\begin{array}{l} 
\widetilde{U}\widetilde{\phi}=\mu\widetilde{\psi}\\ \\
\widetilde{V}\widetilde{\psi}=\mu\widetilde{\chi} \\ \\
\widetilde{W}\widetilde{\chi}=\mu\widetilde{\phi} 
\end{array}\right.\qquad
\left\{\begin{array}{l} 
\mbB_3\widetilde{\phi}=\phi \\ \\
\mbB_1\widetilde{\psi}=\psi \\ \\
\mbB_2\widetilde{\chi}=\chi
\end{array}\right.
\]
(it turns out to be more convenient to take the tilded version of the 
spectral problem, i.e. of the equations in the first triple). In components 
the equations of the spectral linear problem take the form:
\[
\left\{\begin{array}{l}
\widetilde{\bu}_k(1+h\widetilde{\bw}_{k-1})(1+h\widetilde{\bv}_{k-1})
\widetilde{\phi}_k+\lambda\widetilde{\phi}_{k-1}=\mu\widetilde{\psi}_k \\ \\
\widetilde{\psi}_k+\lambda^{-1}\widetilde{\bv}_k(1+h\widetilde{\bu}_k)
(1+h\widetilde{\bw}_{k-1})\widetilde{\psi}_{k+1}=
\mu\widetilde{\chi}_k\\ \\
\widetilde{\chi}_k+\lambda^{-1}\widetilde{\bw}_k(1+h\widetilde{\bv}_k)
(1+h\widetilde{\bu}_k)\widetilde{\chi}_{k+1}=
\mu\widetilde{\phi}_k
\end{array}\right.
\]
while the equations of the evolutionary linear problem take the form
\[
\left\{\begin{array}{l}
(1+h\bu_k)(1+h\bv_k)(1+h\bw_k)\widetilde{\phi}_k+h\lambda\widetilde{\phi}_{k-1}
=\phi_k \\ \\ 
(1+h\bv_{k-1})(1+h\bw_{k-1})(1+h\bu_k)\widetilde{\psi}_k+
h\lambda\widetilde{\psi}_{k-1}=\psi_k\\ \\
(1+h\bw_{k-1})(1+h\bu_k)(1+h\bv_k)\widetilde{\chi}_k+
h\lambda\widetilde{\chi}_{k-1}=\chi_k
\end{array}\right.
\]
After the change of variables $(k,t)\mapsto(k,\tau)$ the spectral equations 
take the form:
\[
\left\{\begin{array}{l}
\widetilde{\mbu}_k(1+h\mbv_{k-1})(1+h\mbw_{k-1})\widetilde{\bphi}_k+\lambda
\bphi_{k-1}=\mu\widetilde{\bpsi}_k \\ \\
\widetilde{\bpsi}_k+\lambda^{-1}\widetilde{\mbv}_k(1+h\widetilde{\mbu}_k)
(1+h\mbw_{k-1})\widetilde{\widetilde{\bpsi}}_{k+1}=
\mu\widetilde{\bchi}_k \\ \\
\widetilde{\bchi}_k+\lambda^{-1}\widetilde{\mbw}_k(1+h\widetilde{\mbv}_k)
(1+h\widetilde{\mbu}_k)\widetilde{\widetilde{\bchi}}_{k+1}=
\mu\widetilde{\bphi}_k 
\end{array}\right.
\]
and the evolutionary equations take the form
\begin{equation}\label{edBL2 lax+ aux1}
\left\{\begin{array}{l}
(1+h\mbu_k)(1+h\mbv_k)(1+h\mbw_k)\widetilde{\bphi}_k+h\lambda\bphi_{k-1}=
\bphi_k \\ \\ 
(1+h\undertilde{\mbv_{k-1}})(1+h\undertilde{\mbw_{k-1}})(1+h\mbu_k)
\widetilde{\bpsi}_k+h\lambda\bpsi_{k-1}=\bpsi_k \\ \\
(1+h\undertilde{\mbw_{k-1}})(1+h\mbu_k)(1+h\mbv_k)
\widetilde{\bchi}_k+h\lambda\bchi_{k-1}=\bchi_k
\end{array}\right.
\end{equation}
or, in the matrix notations,
\begin{equation}\label{edBL2 lax+ aux2}
\left\{\begin{array}{l}
\widetilde{\bphi}=\Delta_3^{-1}\Delta_2^{-1}\Delta_1^{-1}\cF\bphi\\ \\
\widetilde{\bpsi}=\Delta_1^{-1}\undertilde{\!\Delta_5^{-1}}
\undertilde{\!\Delta_4^{-1}}\cF\bpsi\\ \\
\widetilde{\bchi}=\Delta_2^{-1}\Delta_1^{-1}\undertilde{\!\Delta_5^{-1}}\cF\bchi
\end{array}\right.
\end{equation}
Now we use the equations of motion (\ref{dBL2 e}) to bring the spectral 
equations into the form
\begin{equation}\label{edBL2 lax+ aux3}
\left\{\begin{array}{l}
\mbu_k(1+h\mbv_k)(1+h\mbw_k)\widetilde{\bphi}_k+\lambda
\bphi_{k-1}=\mu\widetilde{\bpsi}_k \\ \\
\widetilde{\bpsi}_k+\lambda^{-1}\mbv_k(1+h\mbw_k)(1+h\widetilde{\mbu}_{k+1})
\widetilde{\widetilde{\bpsi}}_{k+1}=
\mu\widetilde{\bchi}_k \\ \\
\widetilde{\bchi}_k+\lambda^{-1}\mbw_k(1+h\widetilde{\mbu}_{k+1})
(1+h\widetilde{\mbv}_{k+1})\widetilde{\widetilde{\bchi}}_{k+1}=
\mu\widetilde{\bphi}_k 
\end{array}\right.
\end{equation}
Next, we transform the spectral equations further with the help of the
evolutionary ones. From the first equations in (\ref{edBL2 lax+ aux1}), 
(\ref{edBL2 lax+ aux3}) we derive:
\[
\bphi_k-(1+h\mbv_k)(1+h\mbw_k)\widetilde{\bphi}_k=h\mu\widetilde{\bpsi}_k
\]
and using this in the first equation in (\ref{edBL2 lax+ aux3}), we bring the
latter into the form
\begin{equation}\label{edBL2 lax+ aux4}
\mbu_k(1+h\mbv_k)(1+h\mbw_k)\widetilde{\bphi}_k+h\lambda
(1+h\mbv_{k-1})(1+h\mbw_{k-1})\widetilde{\bphi}_{k-1}=\mu(\widetilde{\bpsi}_k-
h\lambda\widetilde{\bpsi}_{k-1})
\end{equation}
From the second equation in (\ref{edBL2 lax+ aux1}) we find the expression
\[
(1+h\mbw_k)(1+h\widetilde{\mbu}_{k+1})\widetilde{\widetilde{\bpsi}}_{k+1}=
\frac{1}{1+h\mbv_k}(\widetilde{\bpsi}_{k+1}-h\lambda\widetilde{\bpsi}_k)
\]
and plugging this into the second equation in (\ref{edBL2 lax+ aux3}), we find:
\begin{equation}\label{edBL2 lax+ aux5}
\frac{1}{1+h\mbv_k}\Big(\widetilde{\bpsi}_k+\lambda^{-1}\mbv_k\widetilde{\bpsi}_
{k+1}\Big)=\mu\widetilde{\bchi}_k
\end{equation}
Similarly, from the third equation in (\ref{edBL2 lax+ aux1}) we find the 
expression
\[
(1+h\widetilde{\mbu}_{k+1})(1+h\widetilde{\mbv}_{k+1})
\widetilde{\widetilde{\bchi}}_{k+1}=
\frac{1}{1+h\mbw_k}(\widetilde{\bchi}_{k+1}-h\lambda\widetilde{\bchi}_k)
\]
and plugging this into the third equation in (\ref{edBL2 lax+ aux3}), we find:
\begin{equation}\label{edBL2 lax+ aux6}
\frac{1}{1+h\mbw_k}\Big(\widetilde{\bchi}_k+\lambda^{-1}\mbw_k\widetilde{\bchi}_
{k+1}\Big)=\mu\widetilde{\bphi}_k
\end{equation}
Putting the equations (\ref{edBL2 lax+ aux4}), (\ref{edBL2 lax+ aux5}),
(\ref{edBL2 lax+ aux6}) into the matrix notations, we finally find:
\begin{equation}\label{edBL2 lax+ aux7}
\left\{\begin{array}{l}
\cF^{-1}\mbU\Delta_2\Delta_3\widetilde{\bphi}=\mu\widetilde{\bpsi} \\ \\
\Delta_2^{-1}\mbV\widetilde{\bpsi}=\mu\widetilde{\bchi}\\ \\
\Delta_3^{-1}\mbW\widetilde{\bchi}=\mu\widetilde{\bphi}
\end{array}\right.
\end{equation}
The compatibility conditions of (\ref{edBL2 lax+ aux2}), (\ref{edBL2 lax+ aux7})
coincide with (\ref{edBL2 Lax+ triad}). \qed

\subsection{Explicit dBL2 $\to$ ''relativistic'' BL2}
Our general philosophy and the previous theorem suggest the matrix 
\begin{equation}\label{edBL2 Lax}
\mbL_+=\cF^{-1}\mbU\Delta_2\mbW\Delta_2^{-1}\mbV
\end{equation}
as the Lax matrix of the ''relativistic'' BL2 hierarchy. The spirit
of \cite{GK1}, however, would suggest 
\begin{equation}\label{RBL2 Lax}
\mbL_+=\cF^{-1}\mbU\mbW\mbV
\end{equation}
and this is the choice we actually follow in the next Theorem. However, there
is no conflict between these two suggestions, because changing
\[
\mbW\to\Delta_2\mbW\Delta_2^{-1}
\]
amounts solely to the simple change of variables
\begin{equation}\label{RBL2 to edBL2}
\mbw_k\to\mbw_k\frac{1+h\mbv_k}{1+h\mbv_{k+1}}
\end{equation}
(not touching $\mbu_k$, $\mbv_k$). In other words, in order to make the 
explicit discretization (\ref{dBL2 e}) belonging to the ''relativistic'' BL2 
hierarchy defined below, one needs to perform the change of variables 
(\ref{RBL2 to edBL2}). It turns out that the continuous time
flows look more symmetric, if the Lax matrix is choosen as in (\ref{RBL2 Lax}),
and the discrete time equations (\ref{RBL2 to edBL2}) look
more symmetric by the choice (\ref{edBL2 Lax}) for the Lax matrix.
\begin{theorem}
The hierarchy associated with the Lax matrix {\rm(\ref{RBL2 Lax})}
is described by the following Lax triads:
\begin{equation}\label{RBL2 Lax triads+}
\dot{\mbU}=\cA_1\mbU-\mbU\cA_3,\qquad \dot{\mbV}=\cA_2\mbV-\mbV\cA_0,
\qquad \dot{\mbW}=\cA_3\mbW-\mbW\cA_2
\end{equation}
which have to be supplemented by the identity
\begin{equation}\label{RBL2 Lax triads+ add}
\dot{\cF}=0=\cA_1\cF-\cF\cA_0
\end{equation}
so that for the matrix $\mbL_+=\cF^{-1}\mbU\mbW\mbV$ there holds the Lax 
equation
\begin{equation}\label{RBL2 Lax+}
\dot{\mbL}_+=\Big[\cA_0,\mbL_+\Big]
\end{equation}
Here 
\begin{equation}\label{RBL2 A}
\cA_0=\pi_-(f(\mbL_+))-\sigma(f(\mbL_+))
\end{equation}
and the matrices $\cA_{1,2,3}$ are uniquely defined by the condition of 
compatibility of the equations of the Lax triads {\rm (\ref{RBL2 Lax triads+}),
(\ref{RBL2 Lax triads+ add})}. In particular, for the ''first'' flow of the
hierarchy corresponding to $f(\mbL)=\mbL$, one has:
\begin{eqnarray}
\cA_0 & = & \lambda^{-1}\sum_{k=1}^N(\mbw_{k-1}\mbv_k+\mbu_k\mbv_k
            +\mbu_k\mbw_k+h\mbu_{k-1}\mbw_{k-1}\mbv_k)E_{k,k+1} \nonumber\\
      &   & +\lambda^{-2}\sum_{k=1}^N \mbu_k\mbw_k\mbv_{k+1} E_{k,k+2}
            -h\sum_{k=1}^N\gamma_kE_{kk} \\
\cA_1 & = & \lambda^{-1}\sum_{k=1}^N(\mbw_{k-1}\mbv_k+\mbu_k\mbv_k
            +\mbu_k\mbw_k+h\mbu_k\mbw_k\mbv_{k+1})E_{k,k+1}\nonumber\\
      &    & +\lambda^{-2}\sum_{k=1}^N \mbu_k\mbw_k\mbv_{k+1} E_{k,k+2}
             -h\sum_{k=1}^N\gamma_{k-1}E_{kk} \\
\cA_2 & = & \lambda^{-1}\sum_{k=1}^N(\mbu_k\mbw_k+\mbv_k\mbw_k
            +\mbv_k\mbu_{k+1}+h\mbu_k\mbv_k\mbw_k)E_{k,k+1}\nonumber\\
      &   & +\lambda^{-2}\sum_{k=1}^N \mbv_k\mbu_{k+1}\mbw_{k+1} E_{k,k+2}
            -h\sum_{k=1}^N\gamma_kE_{kk} \\
\cA_3 & = & \lambda^{-1}\sum_{k=1}^N(\mbv_k\mbu_{k+1}+\mbw_k\mbu_{k+1}
            +\mbw_k\mbv_{k+1}+h\mbw_k\mbu_{k+1}\mbv_{k+1})E_{k,k+1}\nonumber\\
      &   & +\lambda^{-2}\sum_{k=1}^N \mbw_k\mbv_{k+1}\mbu_{k+2} E_{k,k+2}
            -h\sum_{k=1}^N\gamma_kE_{kk}
\end{eqnarray}
where
\begin{equation}\label{RBL2 gamma}
\gamma_k=\mbw_{k-1}\mbv_k+\mbu_k\mbv_k+\mbu_k\mbw_k+
h\mbu_{k-1}\mbw_{k-1}\mbv_k+h\mbu_k\mbw_k\mbv_{k+1}
\end{equation}
The equations of motion of the ''first'' flow read:
\begin{eqnarray}\label{RBL2}
\dot{\mbu}_k & = & \mbu_k(\mbv_k+\mbw_k-\mbv_{k-1}-\mbw_{k-1}
+h\mbw_k\mbv_{k+1}-h\mbw_{k-1}\mbv_k+h\gamma_k-h\gamma_{k-1}) \nonumber\\
\dot{\mbv}_k & = & \mbv_k(\mbw_k+\mbu_{k+1}-\mbw_{k-1}-\mbu_k
+h\mbu_k\mbw_k-h\mbu_{k-1}\mbw_{k-1}+h\gamma_{k+1}-h\gamma_k) \\
\dot{\mbw}_k & = & \mbw_k(\mbu_{k+1}+\mbv_{k+1}-\mbu_k-\mbv_k
+h\mbu_{k+1}\mbv_{k+1}-h\mbu_k\mbv_k+h\gamma_{k+1}-h\gamma_k) \nonumber
\end{eqnarray}
\end{theorem}
{\bf Proof} is analogous to the proof of Theorem \ref{RVL hierarchy+}. 
We define $\cA_0$ by (\ref{RBL2 A}) in accordance with the prescription 
of \cite{GK1}, then put $\cA_1=\cF\cA_0\cF^{-1}\in\g_0\oplus\g_-$. The 
matrices $\cA_{2,3}\in\g_0\oplus\g_-$ are uniquely defined by the condition 
of consistency of the Lax triads in (\ref{RBL2 Lax triads+}). The results
for the ''first'' flow (\ref{RBL2}) (= ''relativistic'' BL2) are merely
the specialization of this construction for $f(\mbL)=\mbL$ and follow by
a direct calculation. \qed 
\vspace{1.5mm}

The previous theorem allows to find the equations of motion of all flows
of the hierarchy, and to calculate its integrals. Unfortunately, we still
do not know its Hamiltonian structure, so that the problem of its
integrability in the Liouville--Arnold sense remains open.
%%%%%%%%%%%%%%%%%%%%%%%%%%%%%%%%%%%%%%%%%%%%%%%%%%%%%%%%%%%%%%%%%%%%%%%%%

\section{Conclusion}
We discussed in this paper an approach to finding new integrable hierarchies
based on the time--discretization. Namely, performing the
discretization of a known integrable lattice system and then changing
the Cauchy problem for the discrete time equation, we arrive at another
discrete time equation which belongs, generally speaking, to an integrable
hierarchy disctinct from the one we started with. This new hierarchy is
a regular deformation of the initial one, the deformation parameter being
the time--step of the discretization. 

In the Toda lattice case the new hierarchy turns out to be the relativistic
Toda one, the deformation parameter aquiring the meaning of the inverse
speed of light. This suggests to call in general the new hierarchies
obtained by this procedure ''relativistic'' ones. It would be interesting
to find out, whether this notation allows some physical justification.

In any case, we have now a program of finding ''relativistic'' deformations
of a large number of known lattice hierarchies. In this connection the 
collection of local discretizations in \cite{S6} may serve as a starting
point. For the systems belonging to the lattice KP hierarchy, another
procedure of finding ''relativistic'' deformations was suggested in \cite{GK1},
which has {\it a priori} nothing in common with our one. However, in the
examples of the present paper, both procedures lead to similar results.
It would be important to find out whether there are some deeper principles
behind this coincidence.

\section{Acknowledgements}
Y.S. gratefully acknowledges the stimulating correspondence with Boris
Kupershmidt. On the early stage of this work he also benefited from
discussions with Frank Nijhoff.
%%%%%%%%%%%%%%%%%%%%%%%%%%%%%%%%%%%%%%%%%%%%%%%%%%%%%%%%%%%%%%%%%%%%%%%%%%

\end{document}